\documentclass[%
 aip,
 jcp,
 amsmath,amssymb,
 reprint,%
citeautoscript
]{revtex4-2}

\usepackage{hyperref}
\usepackage{graphicx}
\usepackage{dcolumn}
\usepackage{bm}
\usepackage[utf8]{inputenc}
\usepackage[T1]{fontenc}
\usepackage{mathptmx}
\usepackage{tikz}
\usetikzlibrary{positioning,calc}
\usetikzlibrary{trees}
\usetikzlibrary{decorations.pathmorphing}
\usetikzlibrary{decorations.markings}
\tikzset{fontscale/.style = {font=\relsize{#1}}    }
\usepackage[version=3]{mhchem}
\usepackage{color}
\definecolor{mma1}{rgb}{0.3725,0.5098,0.7020}
\definecolor{mma2}{rgb}{0.8745,0.6078,0.2039}
\definecolor{mma3}{rgb}{0.507813,0.714844,0.2039}
\definecolor{mma4}{rgb}{0.9137,0.3882,0.2398}

\begin{document}

\title{
Stochastic scattering theory for excitation induced dephasing: Time-dependent nonlinear coherent exciton lineshapes
} 

\author{Ajay~Ram~Srimath~Kandada}
\affiliation{Department of Physics and Center for Functional Materials, Wake Forest University, 
1834 Wake Forest Road, Winston-Salem, North Carolina~27109, United~States}

\author{Hao~Li}
\thanks{HL and ARSK contributed equally and are first co-authors of this article.}
\affiliation{Department of Chemistry, University of Houston, Houston, Texas~77204, United~States}

\author{F\'elix~Thouin}
\affiliation{School of Physics, Georgia Institute of Technology, 837 State Street, Atlanta, Georgia~30332, United~States}

\author{Eric~R.~Bittner}
\email{ebittner@central.uh.edu}
\affiliation{Department of Chemistry, University of Houston, Houston, Texas~77204, United~States}

\author{Carlos~Silva}
\email{carlos.silva@gatech.edu}
\affiliation{School of Physics, Georgia Institute of Technology, 837 State Street, Atlanta, Georgia~30332, United~States}
\affiliation{School of Chemistry and Biochemistry, Georgia Institute of Technology, 901 Atlantic Drive, Atlanta, Georgia~30332, United~States}
\affiliation{School of Materials Science and Engineering, Georgia Institute of Technology, North Avenue, Atlanta, Georgia~30332, United~States}

\begin{abstract}

We develop a stochastic theory that treats time-dependent exciton-exciton $s$-wave scattering and that accounts for dynamic Coulomb screening, which we describe within a mean-field limit. With this theory, we model excitation-induced dephasing effects on time-resolved two-dimensional coherent optical lineshapes and we identify a number of features that can be attributed to the many-body dynamics occurring in the background of the exciton, including dynamic line narrowing, mixing of real and imaginary spectral components, and multi-quantum states. We test the model by means of multidimensional coherent spectroscopy on a two-dimensional metal-halide semiconductor that hosts tightly bound excitons and biexcitons that feature strong polaronic character. We find that the exciton nonlinear coherent lineshape reflects many-body correlations that give rise to excitation-induced dephasing. Furthermore, we observe that the exciton lineshape evolves with population time over time windows in which the population itself is static, in a manner that reveals the evolution of the multi-exciton many-body couplings. Specifically, the dephasing dynamics slow down with time, at a rate that is governed by the strength of exciton many-body interactions and on the dynamic Coulomb screening potential. The real part of the coherent optical lineshape displays strong dispersive character at zero time, which transforms to an absorptive lineshape on the dissipation timescale of excitation-induced dephasing effects, while the imaginary part displays converse behavior. Our microscopic theoretical approach is sufficiently flexible to allow for a wide exploration of how system-bath dynamics contribute to linear and non-linear time-resolved spectral behavior. 
\end{abstract}

\date{\today}

\maketitle

\section{Introduction}

\begin{figure*}
    \centering
    \includegraphics[width=16cm]{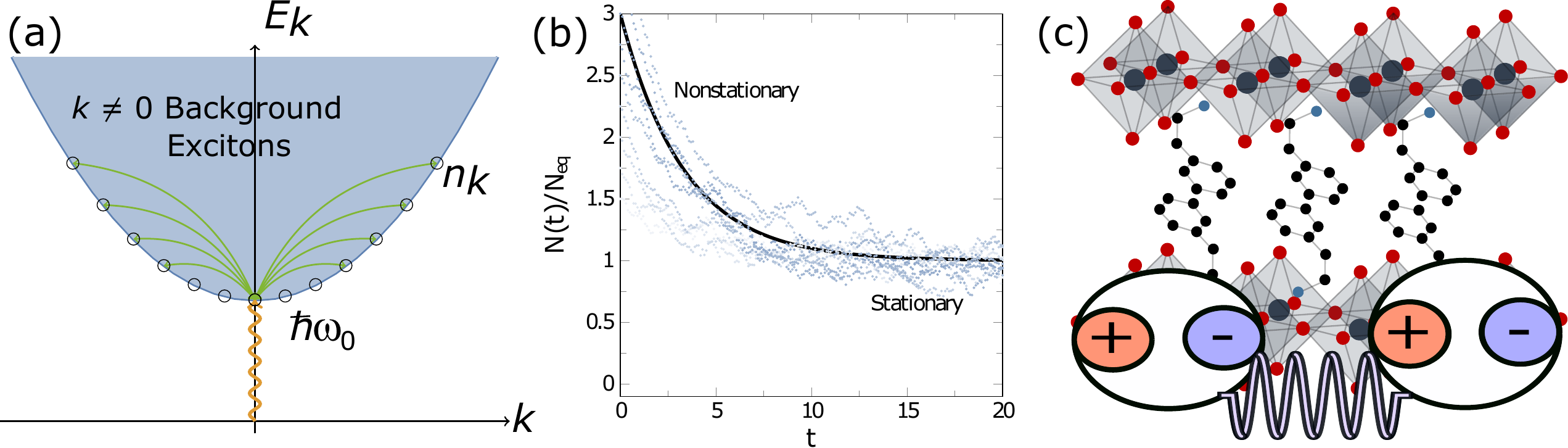}
    \caption{(a) Schematic representation of optical absorption of excitons and exciton-exciton scattering with a background population, where the dispersion relation is in the exciton representation and $\vec{k} = \vec{k_e} + \vec{k_h}$ is the exciton wavevector. (b) Time evolution of population $N(t)/N_{\mathrm{eq}}$ from an initial nonstationary state produced by exciton injection. Individual trajectories are represented by blue dots. Asymptotically, the function reaches a stationary state that yields the Anderson-Kubo limit. (c) Crystal structure of \ce{(PEA)2PbI4} with schematic representation of Coulomb-correlated exciton-exciton elastic scattering.}
    \label{fig:scat}
\end{figure*}

It is well recognized that many-body phenomena
have a profound effect on the linear and non-linear 
optical lineshapes of semiconductors with reduced dimensionality, in which Coulomb correlations can be particularly strong due to decreased screening and quantum confinement effects. One such effect is biexciton formation, in which Coulomb binding of two electron-hole pairs results in new two-electron, two-hole quasiparticles~\cite{mysyrowicz1968excitonic,magde1970exciton,grun1970luminescence,miller1982biexcitons,kleinman1983binding,hu1990biexcitons,brunner1994sharp,albrecht1996disorder,stone2009two,stone2009two,karaiskaj2010two,turner2010coherent}. Another important process that is highly relevant in exciton quantum dynamics is excitation induced dephasing (EID)~\cite{Schultheis1986,Honold1989,Wang1993,Wang1994,Hu1994,Rappen1994,Wagner1997,Wagner1999,shacklette2002role,shacklette2003nonperturbative,Li2006,Moody2011,Nardin2014,moody2015intrinsic,martin2018encapsulation,thouin2019enhanced}, primarily investigated in two-dimensional (2D) systems such as III-V quantum wells~\cite{Honold1989,Wagner1997,Wagner1999,shacklette2002role,Li2006,Moody2011,Nardin2014}, single-layer transition-metal dichalchogenides~\cite{moody2015intrinsic,martin2018encapsulation}, and two-dimensional metal-halide perovskite derivatives~\cite{thouin2019enhanced}. This can be described as the incoherent Coulomb elastic scattering between multiple excitons or between excitons and an electron-hole plasma generated with the excitation optical field. The scattering process gives rise to faster dephasing dynamics compared to the low-density pure-dephasing limit, and may be the dominant dephasing pathway at sufficiently high densities. In many systems, especially those with strong exciton-phonon coupling, the background excitations are transient and co-evolve with optical modes of the system and consequently a strictly incoherent kinetic description such as this mesoscopic approach or a kinetic Markovian Boltzmann-like scattering theory~\cite{Wang1994} cannot describe coherence dynamics. 
EID can be effectively rationalized from a mesoscopic perspective by means of the optical Bloch equations, which capture the effect of many-body exciton scattering on both population and coherence dynamics derived from coherent spectroscopy of semiconductors~\cite{shacklette2002role,shacklette2003nonperturbative}. 
Recent advances towards a more microscopic perspective has been presented by  Katsch et al., in which excitonic Heisenberg equations of motion are used to describe linear excitation line broadening in two-dimensional transition-metal dichalchogenides~\cite{Katsch2020}.
Their results 
indicate exciton-exciton scattering from 
a dark background as a dominant 
mechanism in the power-dependent broadening EID
and sideband formation. 
 Similar theoretical modelling on this class of materials and their van der Waals bilayers have yielded insight into the role of effective mass asymmetry on EID processes~\cite{Erkensten_EID_2020}. These modelling works highlight the need for microscopic approaches to understand nonlinear quantum dynamics of complex 2D semiconductors, but the computational expense could become considerable if other many-body details such as polaronic effects are to be included~\cite{SrimathKandada2020}. As an alternative general approach, an analytical theory of dephasing in the same vein as Anderson-Kubo lineshape theory~\cite{w1954mathematical,kubo1954note}, but that includes
{\em transient} EID and Coulomb screening effects, would be valuable to extract microscopic detail on screened exciton-exciton scattering from time-dependent nonlinear coherent ultrafast spectroscopy, via direct and unambiguous measurement of the homogeneous excitation linewidth~\cite{siemens2010resonance,bristow2011separating}.

Here we employ  a quantum stochastic approach, derived
from a first-principles many-body theory of 
interacting excitons, to develop an analytical model that describes  linear and nonlinear spectral lineshapes that result from exciton-exciton scattering processes,
and, importantly, their dependence on population time due to the 
evolution of a non-stationary/non-equilibrium excitation background (see Fig.~\ref{fig:scat}(a)). 

Our approach is
similar in spirit to the celebrated Anderson-Kubo theory~\cite{w1954mathematical,kubo1954note} and reduces to  
that in the limit of a
stationary background population at sufficiently long times~\cite{paper2}. 
The model captures a microscopic picture of EID by integrating over the 
interactions of excitons produced via a well-defined coherent 
pathway (Fig.~\ref{fig:feynmann} below), and background excitons 
that do not have a well-defined phase relationship that is 
induced by the optical field, and by treating them as a non-stationary
source of quantum noise. In doing so, we can 
directly insert the spectral density of the 
bath into non-linear spectral response functions
and obtain fully analytical expressions for the 
coherent exciton lineshapes. 

We implement the model to investigate the evolution of the two-dimensional coherent excitation lineshape in a polycrystalline thin film of a prototypical two-dimensional single-layer metal-halide perovskite derivative, phenylethylammonium lead iodide [\ce{(PEA)2PbI4}] (see Fig.~\ref{fig:scat}(c) for the crystal structure). We have selected this material as a model system because of its well-defined exciton lineshape
that we have modeled quantitatively within a Wannier-Mott framework~\cite{Neutzner2018} and because it displays strong many-body phenomena --- strongly bound biexcitons at room temperature~\cite{Thouin2018}, and robust EID effects~\cite{thouin2019enhanced}.  Furthermore, we have concluded that the primary excitations are exciton polarons~\cite{thouin2019phonon,SrimathKandada2020} --- quasiparticles with Coulomb correlations that are renormalized by lattice dynamics via polaronic effects; both electron-hole and photocarrier-lattice correlations are ingredients of the system Hamiltonian such that the lattice dressing constitutes an integral component of its eigenstates and eigenvalues. This renders the system highly dynamically disordered such that lattice screening effects play an important role in shaping the linewidth~\cite{thouin2019enhanced} and in dictating nonadiabatic dynamics~\cite{thouin2019polaron}. We measure the dephasing dynamics via the homogeneous linewidth extracted by means of two-dimensional coherent excitation spectroscopy~\cite{siemens2010resonance,bristow2011separating}. In our measurements, excitons generated coherently by a sequence of time-ordered and phase-matched femtosecond pulses scatter from incoherent background excitons and thereby undergo EID, which is perceived via changes of the homogeneous linewidth. We find that EID affects the complex lineshape by mixing absorptive and dispersive features in the real and imaginary spectral components; the real component of the two-dimensional coherent spectrum initially displays a dispersive lineshape that evolves into an absorptive over the timescale in which EID couplings persist, and the imaginary component evolves in the converse fashion. Furthermore, we find that the homogeneous contribution to the spectral linewidth narrows with population time, indicating a dynamic slowing down of the dephasing rate as the EID correlations active at early time dissipate. We find that the dynamic line narrowing phenomenon is reproduced by our stochastic scattering theory, which allows us to explore the effect of dynamic Coulomb screening on EID quantum dynamics.

\section{Theoretical model}

\begin{figure}
    \centering
    \includegraphics[width=2.83cm]{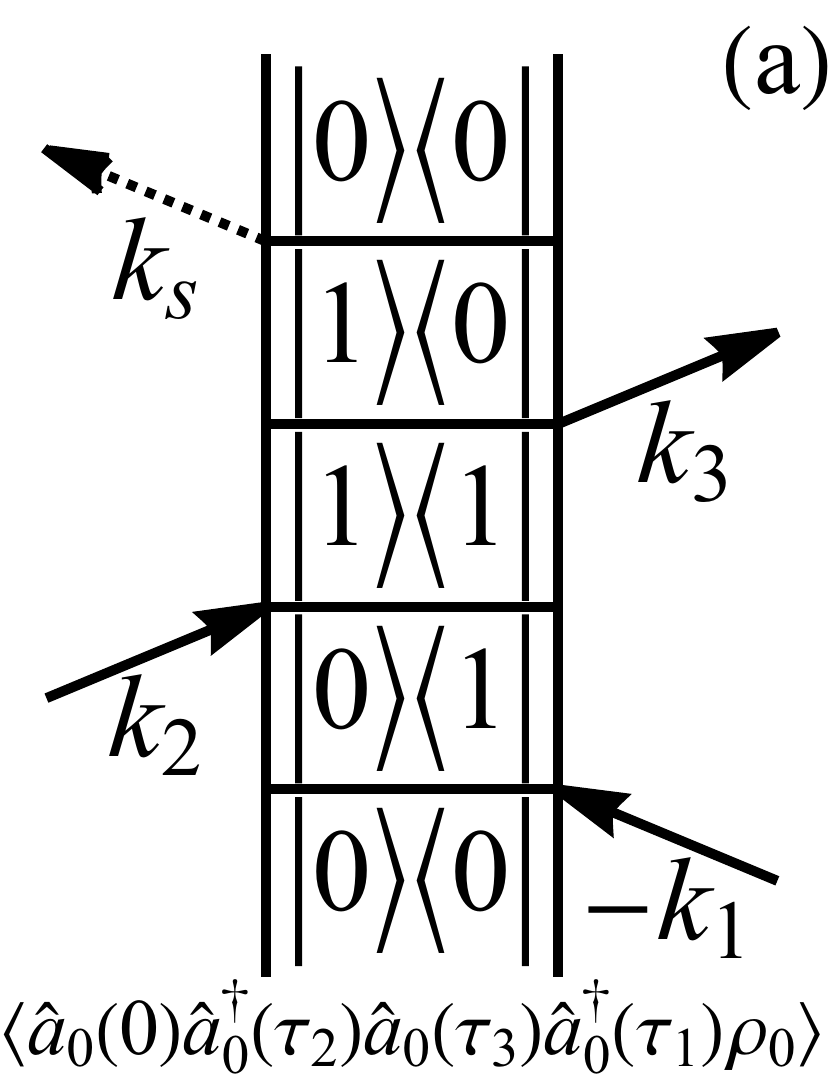}
    \includegraphics[width=2.83cm]{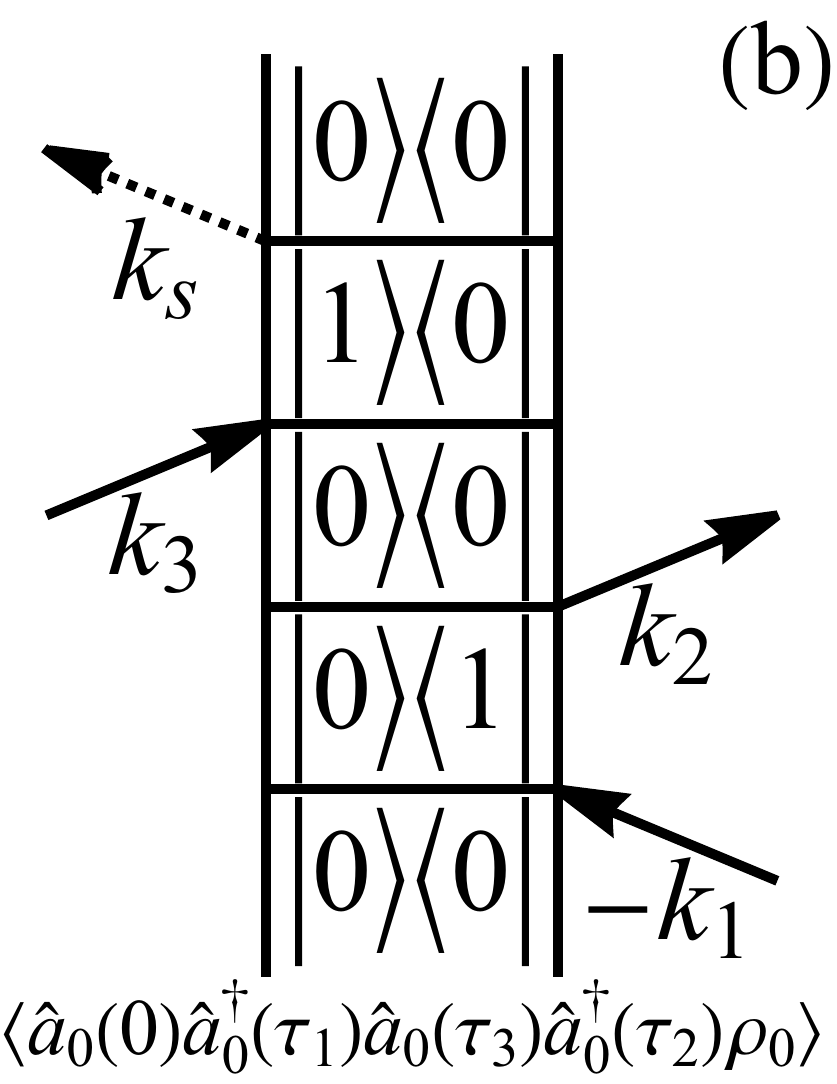}
    \includegraphics[width=2.83cm]{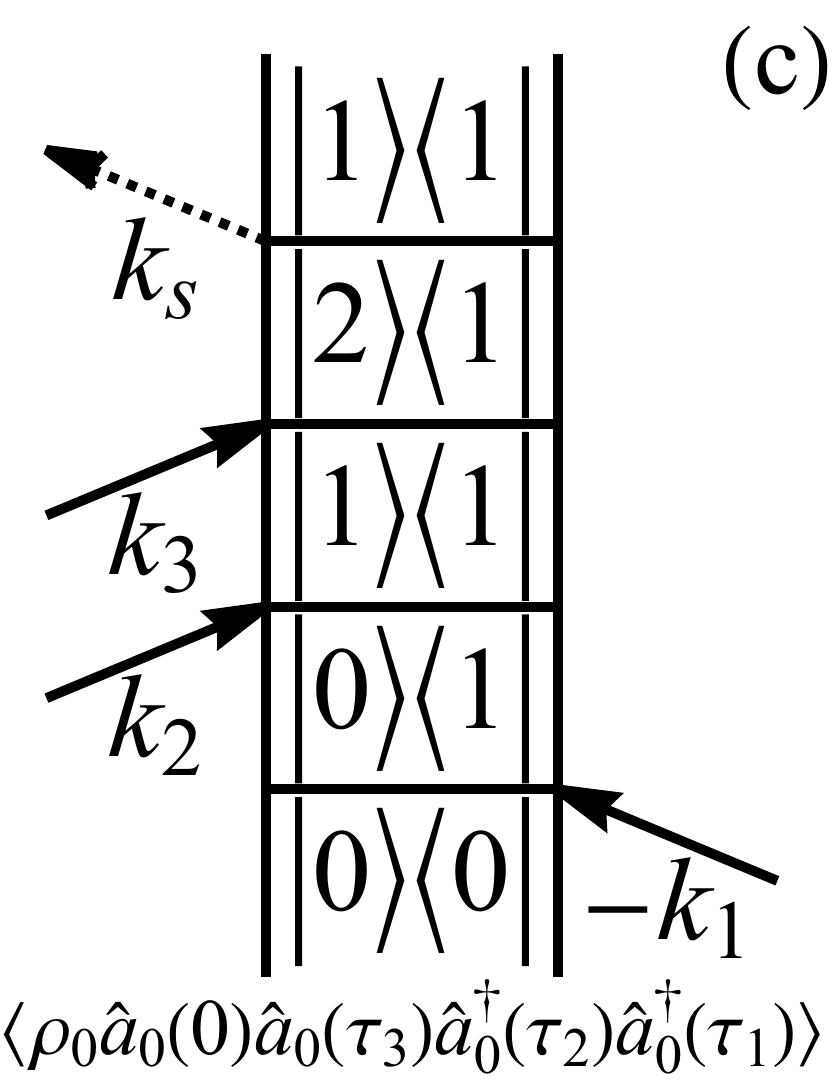}\\
    \includegraphics[width=2.83cm]{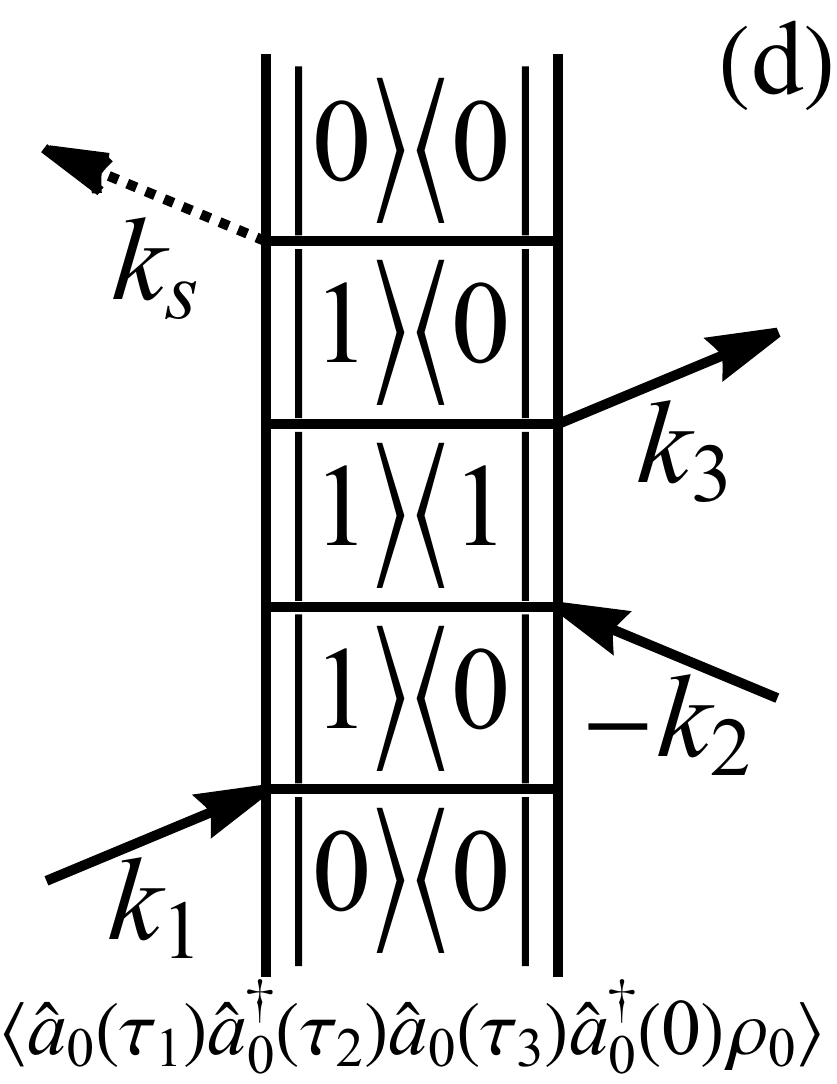}
    \includegraphics[width=2.83cm]{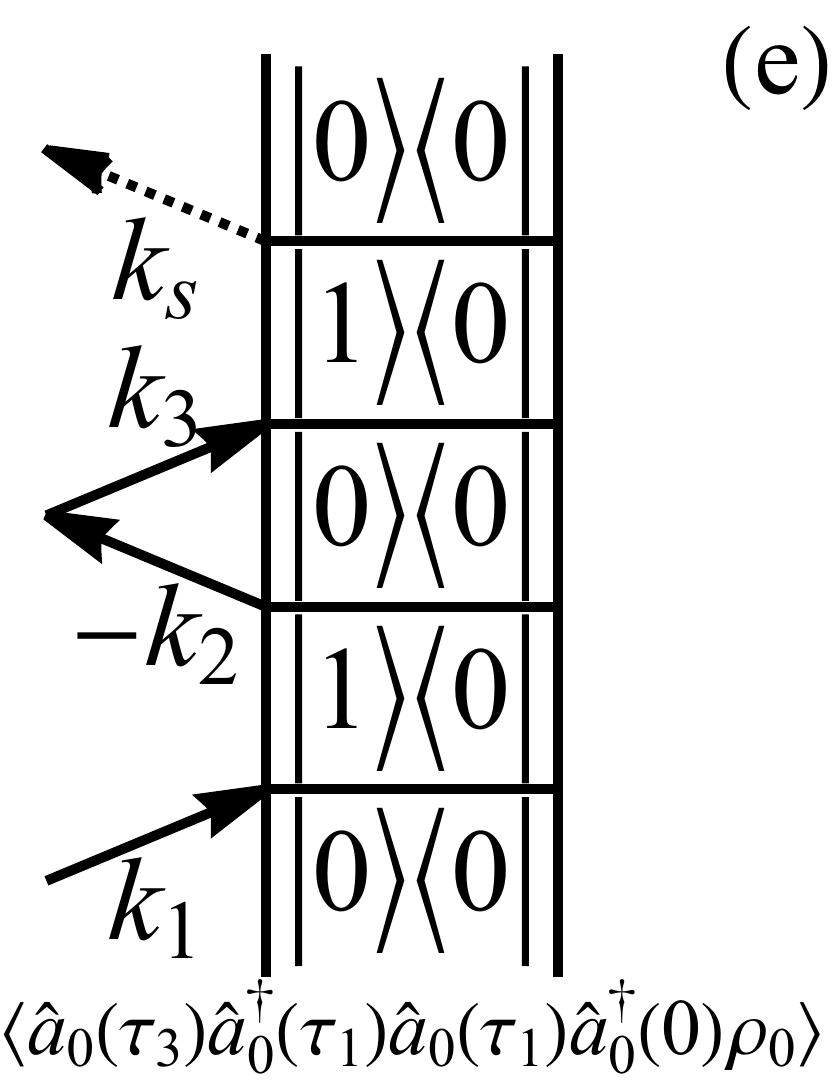}
    \includegraphics[width=2.83cm]{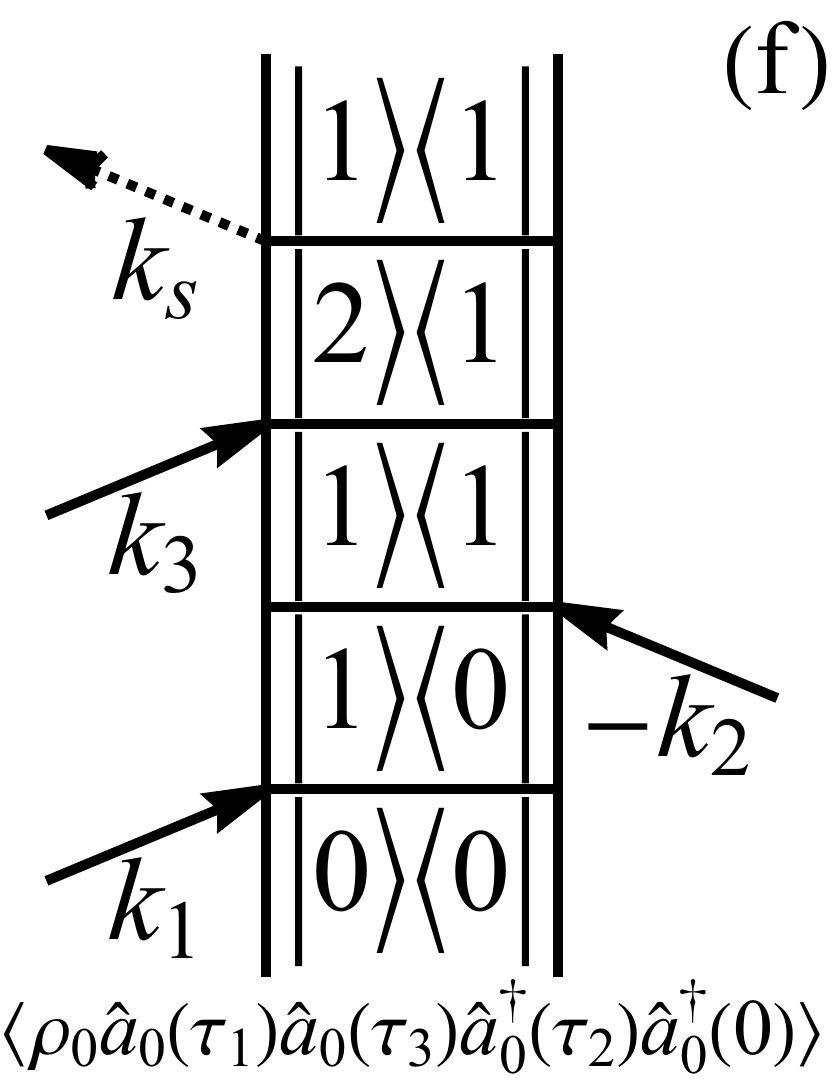}
    \caption{Double-sided Feynman Diagrams for coherent response functions (equation~\ref{eq:Rn}) with rephasing phase matching (top): (a) $R_{2a}$, (b) $R_{3a}$, (c) $R_{1b}^*$, and non-rephasing phase matching (bottom): (d) $R_{1a}$, (e) $R_{4a}$, (f) $R_{2b}^*$. 
    }
    \label{fig:feynmann}
\end{figure}

A complete derivation of our stochastic scattering theory is presented in reference~\citenum{paper2}; here we outline the elements that allow us to calculate nonlinear coherent spectral lineshapes. 
Our model is initiated by assuming that 
at $t=0$ a non-stationary population of background excitations
is created by a broad-band laser excitation.
This physical picture is sketched in Fig.~\ref{fig:scat}. In the current work, excitation occurs with a sequence of phase-matched and time-ordered femtosecond pulses used to measure a coherent nonlinear excitation spectrum, and the excitons produced and measured via a well defined coherent pathway (see Fig.~\ref{fig:feynmann} for the relevant ones in this work) are assumed to scatter elastically with their incoherent counterparts --- excitons that are produced by the pulse sequence but have no phase relationship to those that produse signal in our experiments. 
The initial background population can be 
characterized by an average population $N_0$ and variance $\sigma_{N_0}$ both of which depend upon the 
excitation pulse as well as the density of states of the
material. Optical excitations at $k=0$ evolve 
in concert with a non-stationary ($k\ne 0$) 
background of 
excitations in which the interaction determined by 
a screened Coulomb potential giving rise to a 
noisy driving term that effectively modulates the
exciton energy gap. 
These background excitations undergo diffusion 
as the population relaxes to some stationary distribution.
In reference~\citenum{paper2} we show how one can 
arrive at a reduced model described by a Hamiltonian of the form
 \begin{align}
     H_0(t) \approx \omega_0 \hat{a}^\dagger_0\hat{a}_0 +  \frac{\gamma_1}{2} \hat{a}^\dagger_0 \hat{a}^\dagger_0\hat{a}_0\hat{a}_0 
    +2 \gamma_1 \hat{a}_0^\dagger \hat{a}_0 N(t),
    \label{eq:27}
\end{align} 
whereby $\hat{a}_0$ and $\hat{a}_0^\dagger$ are exciton operators, $N(t)$ is a stochastic variable representing the number of background excitations (see Fig.~\ref{fig:scat}(b) for a depiction of its non-stationary nature), 
$\gamma_1$ is the exciton-exciton interaction, which we obtain from the $s$-wave 
scattering length $a$ and reduced mass $\mu$ within the Born approximation~\cite{born1926quantenmechanik} 
\begin{align}
    \gamma_1 = \frac{4\pi\hbar a}{\mu}.
\end{align}
This assumption does not rely upon the specific form of the exciton-exciton interaction, 
only that it be of finite range. In the current context, this interaction will be due to Coulomb-mediated exciton-exciton scattering that gives rise to EID~\cite{thouin2019enhanced}.  However, it is possible
that each distinct exciton within the family of the 2D perovskite
system considered here~\cite{SrimathKandada2020} have a distinct and unique value of $\gamma_1$, as we reported in ref.~\citenum{thouin2019enhanced}, where we demonstrated distinct Coulomb screening of different exciton polarons. 
For purposes of our theoretical model, we assume that
the system has a {\em single} exciton species that is susceptible to many-body scattering and therefore EID mediated via $\gamma_1$. 
The exciton operators themselves then evolve as 
\begin{align}
    \hat{a}_0(t)&=\exp{\left(-i(\omega_0+\frac{\gamma_1}{2} \hat{n}_0) t - i 2\gamma_1 \int_0^t N(\tau) d\tau \right)} \hat{a}_0, 
    \label{eq:hberg}
\end{align}
where $\hat{n}_0=\hat{a}_0^\dagger\hat{a}_0$ is the number operator of $k=0$ excitons, which is a Schr\"odigner operator and is therefore time independent. This development is within the framework of the interaction representation. 
These operators are then used to construct the
expressions shown schematically in Fig.~\ref{fig:feynmann}. Finally, it should be noted that the angle brackets $\langle \cdots \rangle$ below each diagram denote both the thermal average over the initial conditions as well as averaging over the stochastic variable, $N(t)$. 

A central part of our model is that we assume that the 
background population, 
$N(t) = \langle \sum_{k\ne 0}a_k^\dagger a_k\rangle $ 
follows from an Ornstein-Uhlenbeck process described by 
the stochastic differential equation
\begin{align}
    dN(t) = -\gamma N(t)dt + \sigma dW(t),
\end{align} 
where $W(t)$ represents a Wiener process (continuous-time stochastic process), 
$\gamma$ gives the background
relaxation rate, and $\sigma$ gives the variance.
This is a reasonable assumption 
lacking an explicit description of the background population and
its influence on the system as a source of random collisions. 
Ornstein-Uhlenbeck is a protypical noisy
relaxation process and 
describes the dynamics of an over-damped oscillator driven by 
thermal fluctuations.  

For a stationary 
background population, i.e. $\langle N(t)\rangle  = 0$
the covariance
evolves according to 
$\langle N(t)N(s)\rangle 
= \langle N(t-s)N(0) \rangle 
= \sigma^2 \exp[-\gamma |t-s|]/2\gamma$.
In  this limit, 
our model reduces to the 
Anderson-Kubo  model in which the 
frequency fluctuates about a stationary average
according to an Ornstein-Uhlenbeck process.
In this case, the population relaxation time
in our model is equivalent to the
correlation time in Anderson-Kubo
and  this gives the rate
at which the environment relaxes back to its
stationary average given a small push. Moreover, the
fluctuation amplitude, $\Delta^2$, in Anderson-Kubo
is equivalent to $\omega_0 \gamma_1 \sigma^2/\gamma$ in our model. 
As we shall show, what appears at first to be a 
a simple modification to the
dynamics of a system has rather 
significant implications
in terms of the non-linear 
spectral response of the system.

At time $t=0$, 
we push the background population significantly away
from the steady-state distribution to an initial value of $N(0) = N_0$,
the population evolves as 
\begin{align}
    N(t) = N_0 e^{-\gamma t} + \sigma \int_0^t e^{-\gamma(t-s)}dW(s).
    \label{eq:Nt}
\end{align}
and 
\begin{align}
    \langle N(t)\rangle = e^{-\gamma t}N_0,
\end{align}
where $N_0$ is the mean number 
of background excitations present at time $t=0$.  
In principle, there
will be a distribution about this mean characterized by 
a variance $\sigma^2_{N_0}$.
As a result, we break reversibility and the time symmetry of the 
correlation functions. 
Mathematically, this means 
that 
$\langle N(t)N(s)\rangle \ne \langle N(t-s)N(0) \rangle$
since the choice of initial time is no longer arbitrary.

In Ref.~\citenum{paper2} we discuss the use of It\^o calculus
\cite{Steele2001,VonWeizsacker1990,Fox1987}
to 
evaluate these correlation functions.
From a practical point of view,
the It{\^{o}} calculus is a tool for manipulating  
stochastic processes that are closely related to Brownian motion and
It\^o's lemma allows us to easily perform noise-averaged interactions. 
For the model at hand, 
the covariance of $N(s)$ and $N(t)$ is given by  
\begin{align}
    \nonumber \mathrm{Cov}(N_s,N(t)) =&
     \langle (N_s - \langle N_s\rangle)(N(t) - \langle N(t)\rangle)\rangle \\
&=  \frac{\sigma^2}{2\gamma}\left(e^{-\gamma|t-s|}-e^{-\gamma(t+s)}\right) +  \sigma_{N_0}^2 e^{-\gamma (s+t)},
    \label{eqn:covariance}
\end{align}
 with $\sigma_{N_0}^2$ being the variance of $N(0)$ while the third term vanishes.
 Similarly, the variance
\begin{align}
    \mathrm{Var}[N(t)] = \left(\sigma_{N_0}^2 - \frac{\sigma^2}{2\gamma}\right) e^{-2\gamma t} + \frac{\sigma^2}{2\gamma}
\end{align}
also depends upon the initial variance in the background population.  Mathematically, the Fourier transform of the kernel of the integral in Eq.~\ref{eq:Nt} provides the spectral density of the noisy process.   In fact, a trivial modification of the approach would be to 
replace the kernel in Eq.~\ref{eq:Nt} with another kernel reflecting a more complex spectral density. The resulting expressions for the responses will be more complex indeed.
However, It\^o's lemma provides a tractable route for computing the 
necessary response functions.


\subsection{Optical response functions and  spectral lineshapes}

\begin{figure*}
	\centering
	\includegraphics[width=12.75cm]{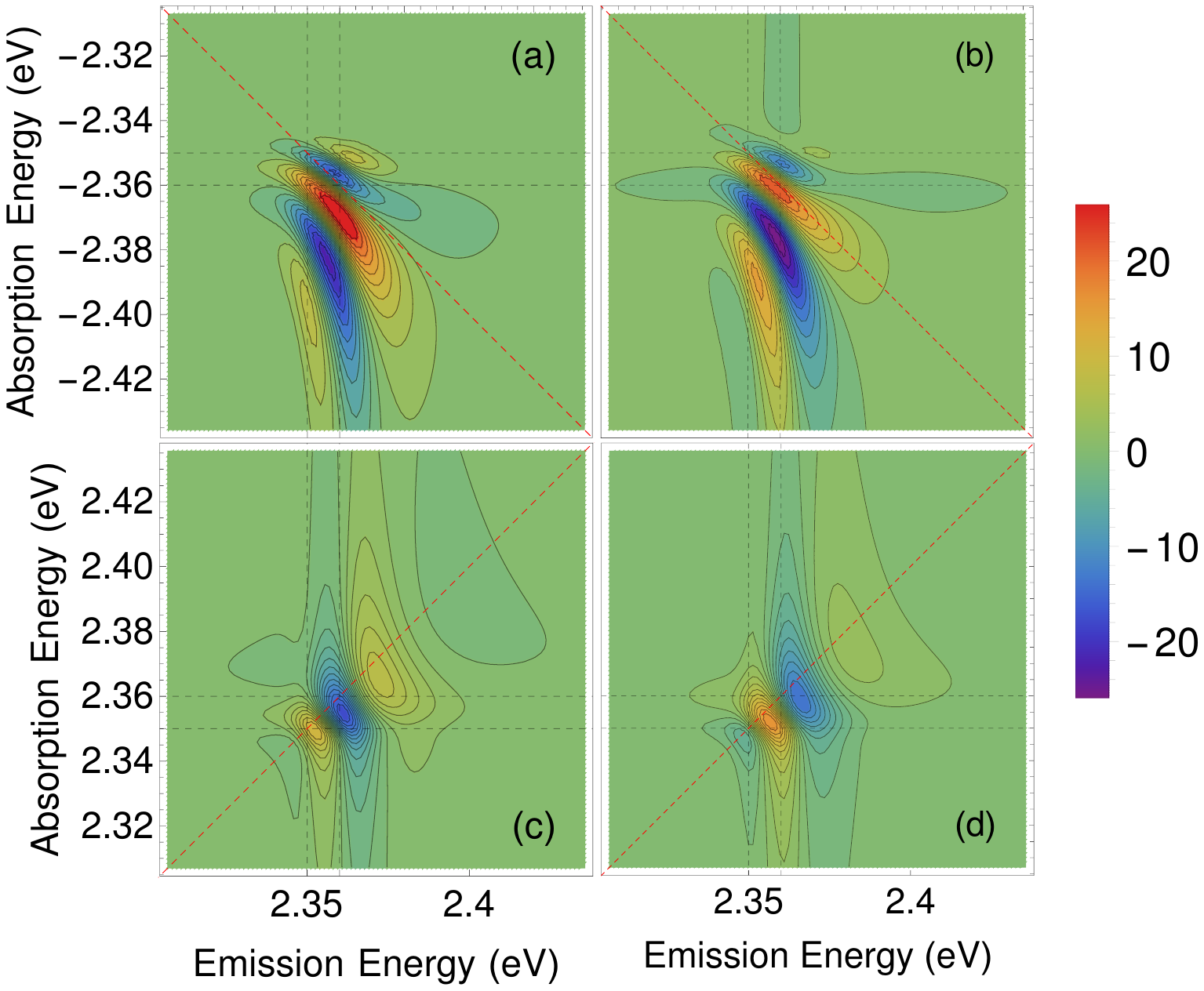}
	\caption{Theoretical real and imaginary spectra, respectively, of rephasing [(a), (b)] and nonrephasing [(c), (d)] phase matching and at population waiting time $\tau_p = 0$\,fs. The vertical false color scale indicated to the right if the figure is in arbitrary units.}
	\label{fig:theoretical_lineshapes}
\end{figure*}

\begin{figure*}
	\centering
	\includegraphics[width=17cm]{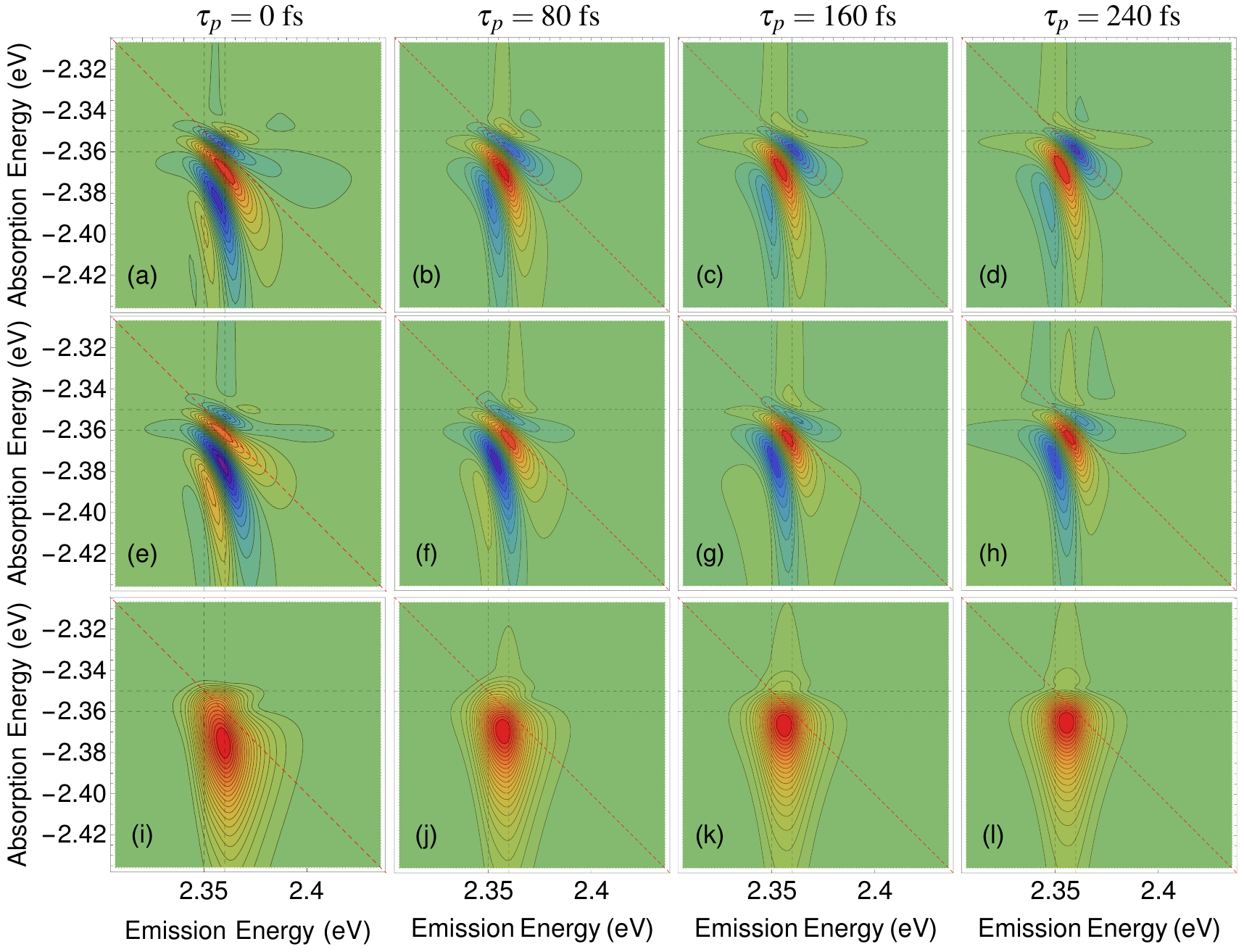}
	\caption{(a)--(d): Real parts of theoretical rephasing spectra at population times $\tau_p$ indicated at the top of each panel. (e)--(h): Corresponding imaginary parts of the spectrum. (i)--(l): The norm (absolute value) of the optical response.}
	\label{fig:theory_time-resolved_spectra}
\end{figure*}

\begin{figure}
	\centering
	\includegraphics[width=8.5cm]{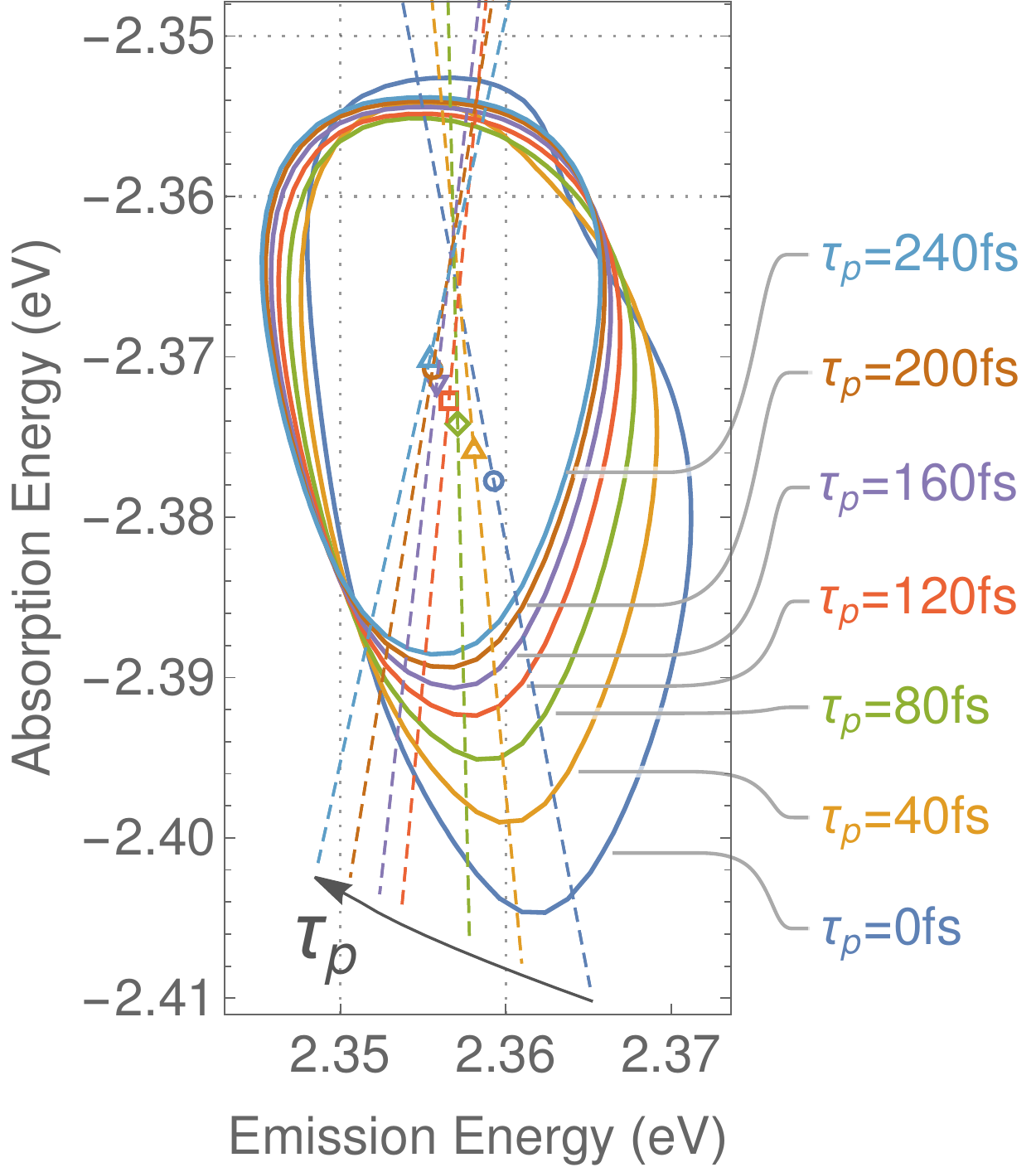}
	\caption{Exciton 2D coherent lineshape contour at half-maximum intensity as a function of population waiting time derived from the theoretical rephasing absolute spectral evolution in Fig.~\ref{fig:theory_time-resolved_spectra}. The center of mass and one of the principle axes are shown for each contour.}
	\label{fig:theory_width}
\end{figure}

Such expressions are useful since they enter directly into the
calculation of response functions for linear and non-linear
spectroscopy. 
For example, the linear response for optical excitation is
given by 
\begin{align}
    S^{(1)}(t) &= \langle \hat\mu(t)[\hat\mu(0),\rho(-\infty)]\rangle,
\end{align}
where $\hat\mu(t) = \mu (\hat a_0^\dagger(t) + \hat a_0(t))$
is the excitonic transition dipole operator and $\rho(-\infty)$ is the initial density matrix.  The absorption spectrum is obtained 
by Fourier transformation.

Averaging over the fluctuations
generates terms involving cumulants of the 
background noise, which result in terms such as 
\begin{align}
    \left\langle \exp\left[ - i \gamma_1 \int_0^tN(\tau) d\tau\right]\right\rangle 
    \approx e^{i\gamma_1 g_1(t) - \frac{\gamma_1^2}{2}g_2(t)},
\end{align}
where $\langle \cdots \rangle$ denotes averaging over noise.  Here, the first cumulant $g_1(t)$ gives rise to a characteristic frequency shift
as the background population decays:
\begin{widetext}
\begin{align}
    g_1(t) = \int_0^t \langle N(\tau) \rangle d\tau = \frac{N_o}{\gamma}(1-e^{-\gamma t}),
\end{align}
and
\begin{align}
  g_2(t,t')=  \nonumber \int_0^t \int_0^{t'} \mathrm{Cov}& \left[N(\tau), N(\tau')\right] d\tau' d\tau = \\
    &\frac{\sigma^2}{2\Gamma^3}\left[2\Gamma \mathrm{min}(t,t') + 2e^{-\Gamma t} + 2e^{-\Gamma t'} - e^{-\Gamma|t'-t|} - e^{-\Gamma(t'+t)} -2\right] \\
    \nonumber &+ \frac{\sigma_{N_o}^2}{\Gamma^2}\left[e^{-\Gamma(t+t')}-e^{-\Gamma t}-e^{-\Gamma t'}+1\right].
\end{align}
When the two time limits are the same, this reduces to
\begin{align}
    g_2(t) &= \int_0^t\int_{0}^t{ \rm Cov}[N(\tau),N(\tau')] d\tau d\tau' \nonumber \\
    &= \frac{\sigma^2}{2\gamma^3}(2\gamma t + 4e^{-\gamma t} -  e^{-2 \gamma t} - 3) 
    +  
    \frac{\sigma_{N_0}^2}{\gamma^2}(1-e^{-\gamma t} )^2.
    \label{eq:kubo-like}
\end{align}

\end{widetext}%
In Ref.~\citenum{paper2} we discussed the linear response of our model and
its relation to the Anderson-Kubo model.  Here we shall focus solely on the 
higher-order responses that reveal the dynamic evolution of the two-dimensional coherent excitation line-shape.
The third-order response 
involves phase-matched interactions of the system with a sequence of three laser pulses: 
\begin{widetext}
\begin{align}
    S^{(3)}(\tau_3,\tau_2,\tau_1) =\langle
    \mu(\tau_3)[\mu(\tau_2),[\mu(\tau_1),[\mu(0),\rho(-\infty)]]]]
    \rangle.
\end{align}
The times $0< \tau_1 < \tau_2 < \tau_3$ define the sequence of the time-ordered 
interactions in Fig.~\ref{fig:feynmann}.
The expressions 
for these can evaluated using the standard rules for 
double-sided Feynman diagrams (Fig.~\ref{fig:feynmann}, c.f.\ Ref.~\citenum{Mukamel1995}) 
representing 
various optical paths that for a given 
pathway take the form 
\begin{align}
    R_n(\tau_1,\tau_2,\tau_3) 
    &=
    -\left(\frac{i}{\hbar}\right)^4\mu^4
    e^{-i(\omega_0 + n_0\gamma_1)(\pm \tau_3 \pm \tau_2 \pm \tau_1})
    \left\langle
    \exp\left[i\gamma_1^2\sum_{j=1}^3(\pm)_j\int_0^{\tau_j} N(s)ds \right]
    \right\rangle \\
    &= 
    n_0^2 \exp\left[-i(\omega_0 + n_0\gamma_1)\sum_{j=1}^3
    (\pm)_j\tau_j\right]\nonumber \\
    &\times\exp[-i \gamma_1 g_1(\pm\tau_1,\pm\tau_2,\pm\tau_3)]
    \exp\left[-\frac{\gamma_1^2}{2}g_2(\pm\tau_1,\pm\tau_2,\pm\tau_3)
    \right].
    \label{eq:Rn}
\end{align}
\end{widetext}
The sign function $(\pm)_j$ takes ``$+$'' and ``$-$'' depending upon the sign of the photon wavevector entering or leaving the system.
Note that $g_{1}(\tau_j)=0$ when the system is initially prepared in the ground state since $N_0=0$. Fig.~\ref{fig:feynmann} shows the most relevant diagrams for the rephasing ($-$) and non-rephasing ($+$) optical response.

It is important to notice that the
the exciton-exciton interaction term $\gamma_1$, 
and hence the screening due to exciton-lattice interactions,
appears in three distinct places in the 
third-order responses.  First, as a  
frequency shift due to  self-interactions between
the bright excitons. Second, as a frequency shift
due to interactions of bright excitons with the 
evolving background population density.  Third, 
as the leading contribution to the 
lineshape. In addition, the third term involving $g_2(t)$  carries the influence of the 
initial conditions (via $\sigma_{N_0}$). The effect of many-body exciton-exciton scattering thus leads to time-evolving EID processes. Given these observations, we expect that the homogeneous linewidth will evolve with population time, dictated by the evolution of $g_2(t)$.

\section{Two-dimensional coherent spectroscopy}

\subsection{Predictions from the stochastic model}\label{sec:theo_pred}

Having established the mathematical model, let us 
briefly recapitulate some of its features.
First, we started by assuming that the background population dynamics give rise to a
stochastic process $N(t)$ that enters into the Heisenberg equations of motion for 
the system operators (Eq.~\ref{eq:hberg}). In particular, we assumed that $N(t)$
corresponds to an overdamped Brownian oscillator and that at time $t=0$ there is a
non-stationary population of background excitations.   
These two mathematical assumptions can be relaxed to some extent if one has a 
more detailed description of the spectral density of the background process and 
the initial background population.  Secondly, we assume that averages over 
exponential terms can be evaluated
using the cumulant expansion. What then follow are the 
mathematical consequences as expressed in terms of the spectral responses of the model. 
In Ref.~\citenum{paper2} we explored the linear response, especially as compared to the
Anderson-Kubo model~\cite{w1954mathematical,kubo1954note}. 
The key features of our model include~\cite{paper2}:
\begin{enumerate}
    \item {\em Blocking}: Increasing the initial background exciton density suppresses the peak 
    absorption intensity. 
     \item {\em Energy shift}: The peak position shifts to the blue with increasing
     background population due to increased Coulombic interactions. 
     \item {\em Broadening:} The spectrum acquires a log tail extending to the 
     blue due to the dynamical evolution of the background.  
     This feature appears in the 2D coherent spectroscopy as an asymmetry
     along the absorption axis and as phase scrambling in the rephasing and non-rephasing 
     signals (see Figs.~S4 and S5 in the Supplemental Material of ref.~\citenum{thouin2019enhanced} for the exciton-density-dependent lineshapes at $\tau_p = 0$).
     \item {\em Biexciton formation:} The peak is split by $\gamma_1/2$ corresponding to the 
     biexciton interaction~\cite{Thouin2018}. 
\end{enumerate}
These effects are consistent with experimental  
observations and theoretical models of 2D semiconductors 
and transition metal dicalcogenides~\cite{Katsch2020}.

Figs.~\ref{fig:theoretical_lineshapes}, \ref{fig:theory_time-resolved_spectra}, and \ref{fig:theory_width}
correspond to the rephasing and non-rephasing behavior of theoretical model as parametrized 
to approximate the excitons in the 2D metal-halide perovskite system studied in the 
experimental investigations, which we shall describe later in this section. The parameters used to produce these spectra are given in Table~\ref{tab:parameters}. 
The two pairs of gray dashed lines correspond to the bare exciton energy at 
$\hbar\omega_0 = 2.35$\,eV and the 
dressed exciton energy at $\hbar\omega_0 + \gamma_1/2  = 2.36$\,eV. 
Fig.~\ref{fig:theoretical_lineshapes} gives the rephasing (a,b) and non-rephasing (c,d) spectra
computed at $\tau_p = 0$.  
Two features highlighted above are immediately striking in the modelled 2D spectra. Both the  asymmetry of the 
signals as well as the lineshape inversion of the real and imaginary spectral components can be traced specifically to 
terms within the response functions in Eq.~\ref{eq:Rn} that depend upon the
transient background relaxation and exciton self-interactions. 

Both the phasing and asymmetry evolve with increasing population time as 
shown in Fig.~\ref{fig:theory_time-resolved_spectra}(a-l).  Importantly, 
the rephasing signal evolves being dispersive at $\tau_p = 0$ to 
absorptive at longer times. The non-rephasing signal [Fig.~\ref{fig:theory_time-resolved_spectra}(e-h)] has complementary behavior, evolving 
from absorptive to dispersive.  
Figs.~\ref{fig:theory_time-resolved_spectra}(i-l)
give the absolute value of the total response as it evolves 
over $\tau_p$. The peak is displaced from the diagonal 
and its position as well as the linewidth evolves over $\tau_p$.  

In Fig.~\ref{fig:theory_width} we extract the contour corresponding to the 
half-maximum intensity at various indicated $\tau_p$ population times.   
Superimposed over each contour is one of the 
principal axes of the contour scaled according to its magnitude. The central points are the
geometric centers of contours.  This analysis clearly shows that the peak systematically
narrows, rotates, and distorts as the exciton co-evolves with the background population.
Moreover, the center peak shifts by about 10\,meV towards the red in both absorption and 
emission spectral dimensions as Coulombic interactions
with the evolving background are diminished --- this phenomenon is known as excitation-induced shift~\cite{karaiskaj2010two}. The lineshape evolution predicted by the stochastic model is due to $g_1(t) \neq 0$. The early-time blue shift as well as more rapid dephasing arise from many-body effects contained within $g_1$; as this function decays these effects dissipate as shown in Fig.~\ref{fig:theory_width}. We note that in equation~\ref{eq:Rn}, if we set $g_1=0$ the coherent response functions reduce to a stationary background, and the lineshape evolution in Fig.~\ref{fig:theory_width} would not arise --- see also reference~\citenum{paper2}.

It should be noted, however, that we assume here that the initial background excitation 
is broad compared to its fluctuations about a stationary state.  Starting from the opposite
regime, one can obtain dynamic broadening (rather than narrowing) as the system relaxes to the 
stationary state. We specifically choose these conditions to best represent the experimental 
conditions of an ultrafast experiment with $\sim 20$\,fs pulses.

\begin{table}[]
    \centering
    \begin{ruledtabular}
    \begin{tabular}{lcl}
         \textbf{Description}          &  \textbf{Symbol}    & \textbf{Value} \\ 
         \hline
         bare exciton energy  & $\hbar\omega_0$     & 2.35\,eV\\
         noise variance       & $\sigma^2$          &     0.0025\,fs$^{-1}$\\
         relaxation rate      & $\gamma$            &    0.01\,fs$^{-1}$ \\
         exciton/exciton interaction & $\gamma_1$   &    20\,meV \\
         avg.\ init.\ background density  & $N_0$        &  2 per unit volume\\
         init.\ background variance       & $\sigma_{N_0}$ & 0.35 per unit volume  \\
    \end{tabular}
    \end{ruledtabular}
    \caption{Parameters used in the theoretical model to produce Figs.~\ref{fig:theoretical_lineshapes}, \ref{fig:theory_time-resolved_spectra}, and \ref{fig:theory_width}.}
    \label{tab:parameters}
\end{table}

In order to test the predictions of the non-stationary model described above, we have carried out two-dimensional coherent measurements on \ce{(PEA)2PbI4} (PEA = pheny\-ethyl\-ammonium) --- a multiple-quantum-well-like single-layer metal-halide perovskite derivative (see Fig.~\ref{fig:scat}(c)). We choose this material to test the theoretical framework developed above because of its susceptibility to strong many-body effects~\cite{Kato2003,Thouin2018,thouin2019enhanced} and dynamic exciton-lattice coupling that drives their dynamics~\cite{Neutzner2018,thouin2019phonon,thouin2019polaron,SrimathKandada2020}. To further examine EID in this material, we dissect the population-time-resolved nonlinear coherent optical lineshape of the family of exciton polarons~\cite{SrimathKandada2020} by means of two-dimensional (2D) coherent spectroscopy~\cite{cho2008coherent}. Importantly for this work, the 2D coherent optical lineshape permits separation of the homogeneous and inhomogeneous contributions to the linewidth~\cite{tokmakoff2000two,siemens2010resonance,bristow2011separating}, and is therefore an appropriate technique to spectrally and temporally resolve dephasing rates, which we exploit here to quantify EID dynamics. We have discussed the linear spectral lineshape of \ce{(PEA)2PbI4} in references~\citenum{Neutzner2018} and \citenum{Thouin2018}, and here we summarize it in section~\ref{sec:linearspec}. We will then discuss the 2D spectral lineshapes of \ce{(PEA)2PbI4} in section~\ref{sec:2Dspec}. Specifically, we will present the experimentally observed dispersive lineshape as a signature of EID, which we had suggested previously in Ref.~\citenum{thouin2019enhanced}. We will show that such a lineshape evolves into absorptive form with population time, as a consequence of $g_2(t)$ in the response function which is quenched due to the increased screening of exciton-exciton interactions.

 \subsection{Linear absorption lineshape}\label{sec:linearspec}
 
 \begin{figure}
 	\includegraphics[width=8.0cm]{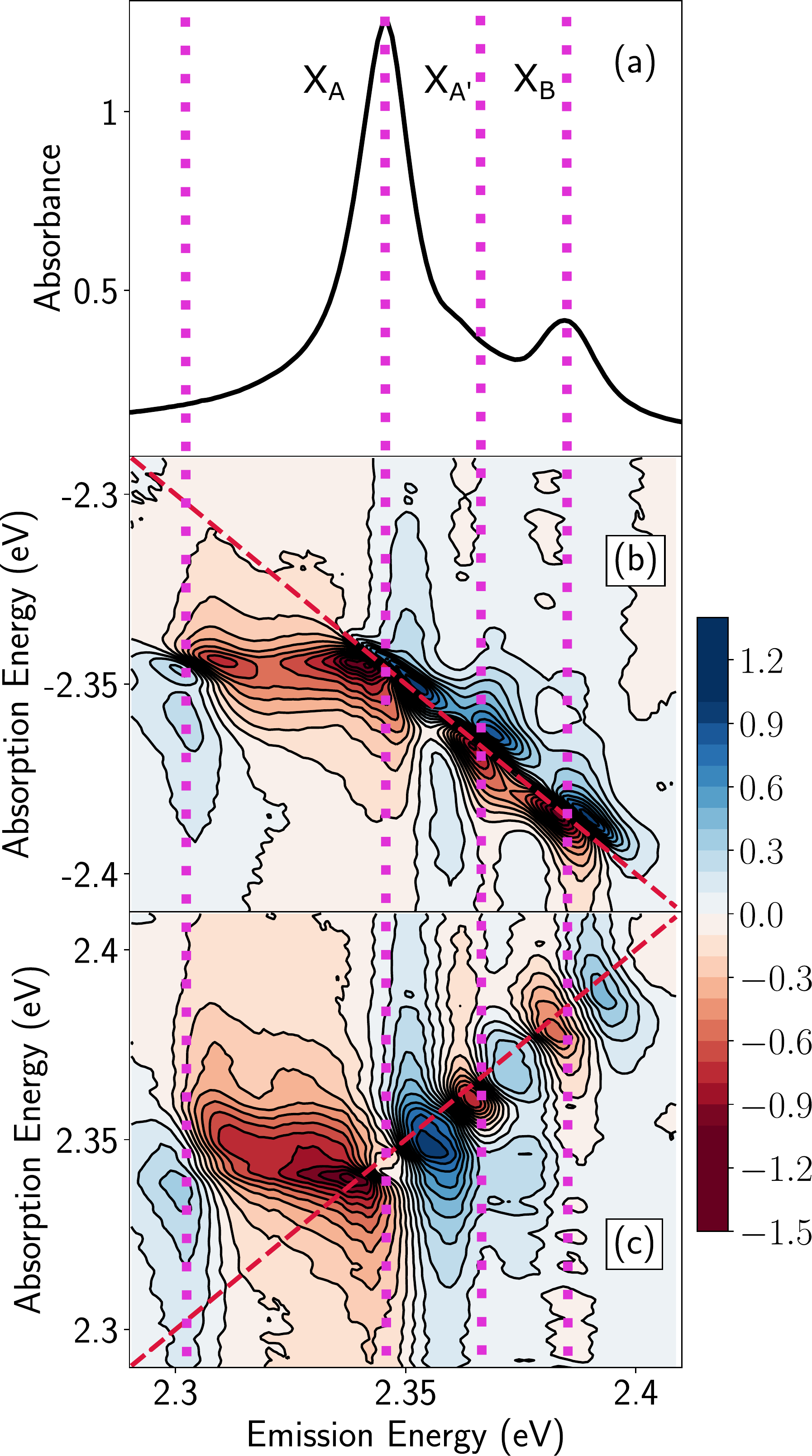}
 	\caption{(a) Linear absorption spectrum of \ce{(PEA)2PbI4} at 5\,K. 
 		Real part of the corresponding rephasing (b) and non-rephasing (c) spectra 
 		at a population time of $\tau_p=0$\,fs. The bar to the right of the figure displays the vertical false color scale in arbitrary units.}
 	\label{fig:spectra1}
 \end{figure}

\begin{figure*}[th]
	\centering
	\includegraphics[width=17cm]{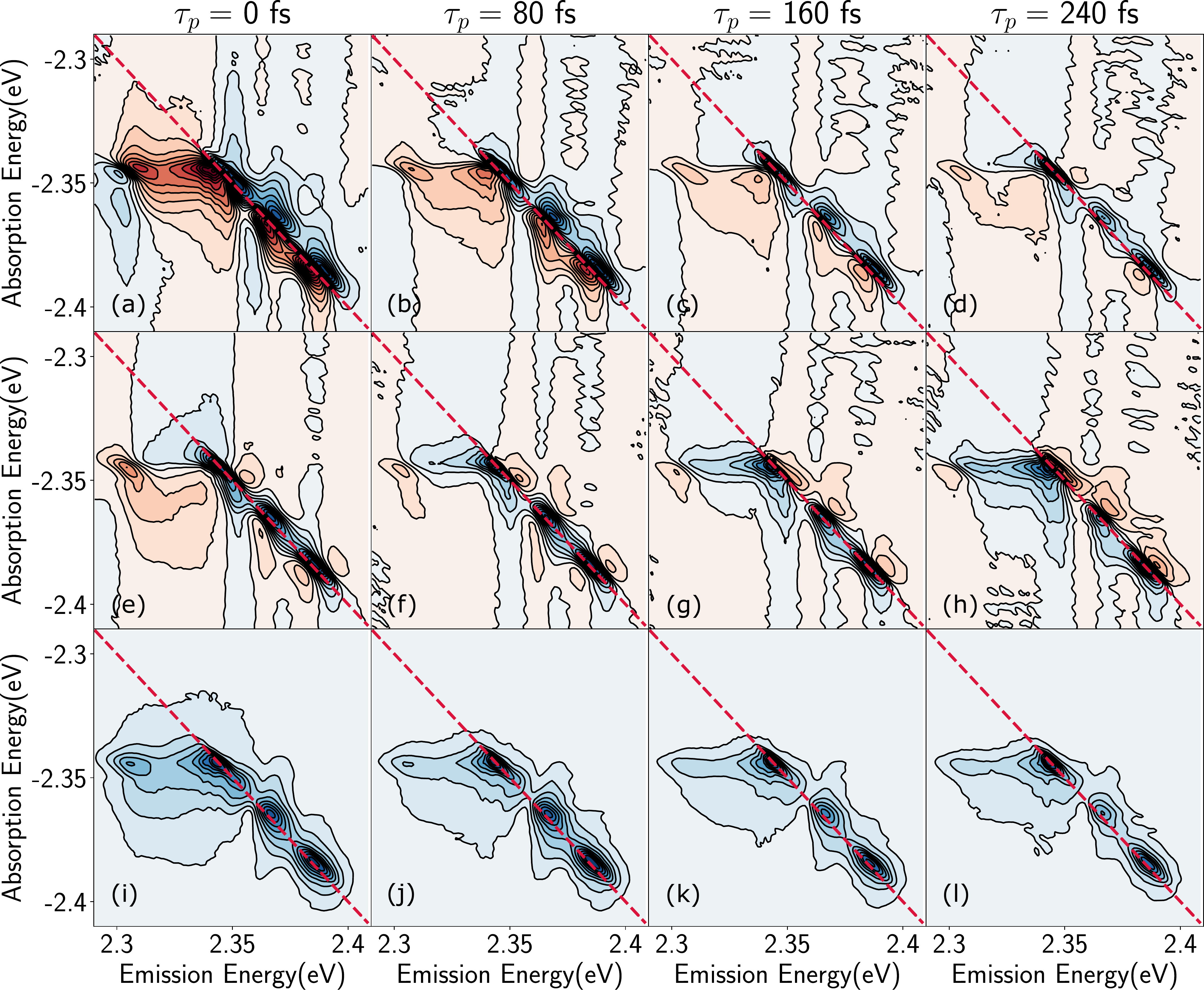} 
	\caption{(a)--(d): Real parts of experimentally measured rephasing spectra at population times $\tau_p$ indicated at the top of each panel, measured at 5\,K. (e)--(h): Corresponding imaginary parts of the spectrum. (i)--(l): The norm (absolute value) of the optical response. All spectra components are plotted in the same relative vertical color scale to facilitate comparison of the time-dependent signal.}
	\label{fig:spectra2}
\end{figure*}

\begin{figure}
	\centering
	\includegraphics[width=8.5cm]{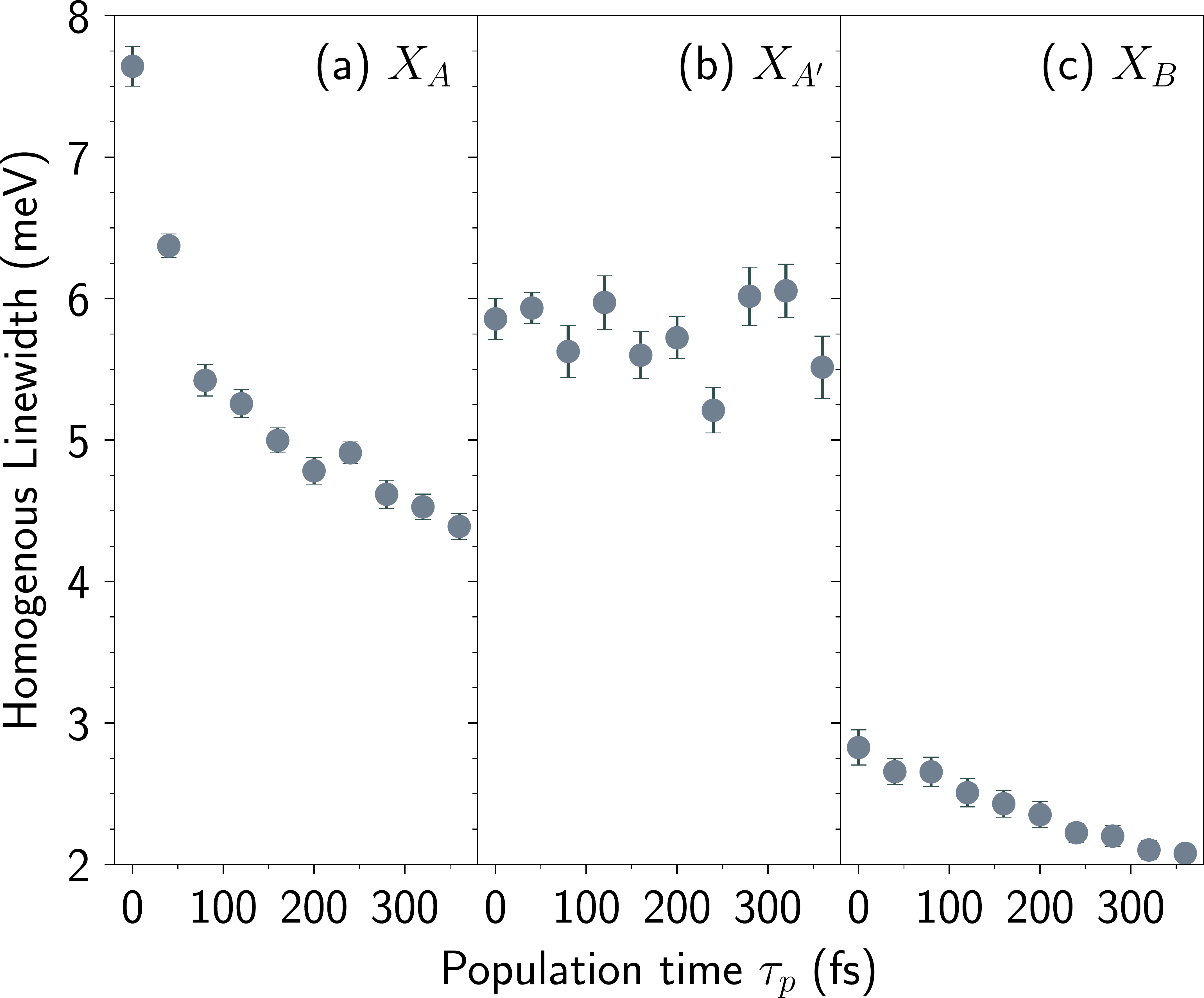}
	\caption{Homogenous linewidths obtained from the lineshape analysis of the absolute value of the rephasing spectra (see reference~\citenum{thouin2019enhanced}) plotted as a function of the population time for (a) $X_A$, (b) $X_{A^{\prime}}$ and (c) $X_B$ exciton lines shown in Fig.~\ref{fig:spectra1}(a).}
	\label{fig:width}
\end{figure}
 
We display the linear absorption spectrum of \ce{(PEA)2PbI4} measured at 5\,K in the region of exciton absorption, in Fig.~\ref{fig:spectra1}(a). It has been reported extensively that this spectrum displays structure consisting of multiple resonances~\cite{gauthron2010optical,Tanaka2002,Tanaka2005,Neutzner2018,Thouin2018,Straus2018a,mauck2019excitons,SrimathKandada2020}. We have argued previously that the primary photoexcitations in two-dimensional hybrid metal-halide perovskite derivatives are a family of exciton polarons~\cite{SrimathKandada2020}, with exciton binding energies differing by $\sim 40$\,meV~\cite{Neutzner2018}, and each with distinct phonon dressing~\cite{thouin2019phonon}. Here, we focus on the two primary transitions labelled $X_A$ and $X_B$, but we  also highlight an additional shoulder of $X_A$, labelled $X_{A^{\prime}}$. We had initially hypothesized that $X_{A^{\prime}}$ is the envelope of replicas in a vibronic progression with origin $X_A$~\cite{Neutzner2018}, but we then subsequently found that its elastic scattering rate is distinct from $X_A$ and $X_B$~\cite{thouin2019enhanced}, indicating it to be another distinct state within the spectral bandwidth of the excitonic transitions. We highlight that this spectral structure is general to other derivatives with different organic cations, including ones that induce lattice distortions~\cite{cortecchia2016broadband} that modulate the central exciton binding energy~\cite{Neutzner2018,tremblay20194npea}, and the relative intensities of the transitions, but not the energy spacing in the spectral structure~\cite{Neutzner2018,SrimathKandada2020}. 

\subsection{Time-resolved 2D coherent excitation lineshape}\label{sec:2Dspec}

We next consider the complex 2D coherent excitation spectrum to quantify the consequences of EID in the nonlinear lineshape. We have previously reported that the multiple excitons identified in Fig.~\ref{fig:spectra1}(a) display strong many-body effects, manifested via the presence of stable biexcitons~\cite{Thouin2018} and the dominance of EID signatures on the homogeneous linewidth~\cite{thouin2019enhanced}. We have observed that $X_A$ and $X_B$ display different dependence of EID on exciton density and on temperature~\cite{thouin2019enhanced} and have interpreted these phenomena as indicative of specific dynamic Coulomb screening of $X_A$ and $X_B$ by different polaronic dressing phonons~\cite{thouin2019phonon}. 

Shown in Figs.~\ref{fig:spectra1}(b) and \ref{fig:spectra1}(c) are the real parts of two different coherent excitation pathways; the time-ordering of the three optical pulses in the experiment and phase-matching conditions define the specific excitation pathways, based on which \textit{rephasing} [Fig.~\ref{fig:spectra1}(b)] and \textit{non-rephasing} [Fig.~\ref{fig:spectra1}(c)] spectra are obtained~\cite{cho2008coherent}. In the rephasing experiment, the pulse sequence is such that the phase evolution of the polarization after the first pulse and the third pulse are of opposite sign, while in the non-rephasing experiment, they are of the same sign (see equation~\ref{eq:Rn} and Fig.~\ref{fig:feynmann}). Both measurements shown in Fig.~\ref{fig:spectra1} are taken at a population waiting time $\tau_{p} = 0$\,fs and an excitation fluence of 40\,nJ/cm$^{-2}$, which corresponds to an exciton density in which we have identified effects of elastic exciton-exciton scattering~\cite{thouin2019enhanced}. Corresponding diagonal spectral features at the energies of $X_A$, $X_{A^{\prime}}$ and $X_B$ (indicated by the magenta vertical dotted-lines in Fig.~\ref{fig:spectra1}) are observed, both in rephasing and non-rephasing spectra. Apart from these diagonal peaks, we observe an off-diagonal excited-state absorption feature (opposite phase with respect to the diagonal features) corresponding to a correlation between the absorption energy of $X_A$ and emission energy $\sim2.3$\,eV, which has no corresponding diagonal signal. We have assigned this cross-peak to a biexciton resonance~\cite{Thouin2018}.

From the norm of the rephasing spectrum at zero time [not shown in Fig.~\ref{fig:spectra1} but shown below in Fig.~\ref{fig:spectra2}(i)], one can extract the homeogeneous and inhomogeneous linewidths via a global analysis of the diagonal and the anti-diagonal lineshape~\cite{siemens2010resonance,bristow2011separating}. In reference~\citenum{thouin2019enhanced} we reported that  the homogeneous linewidth ($\gamma_0$) in the absence of excitation induced dephasing (zero-density limit) is approximately 2\,meV for all the observed excitonic resonances, that implies a dephasing time $T_2^* \sim 500$\,fs.~\footnote{We note that $\gamma$, and not $2\gamma$, as defined in ref.\ 27, is the homegeneous linewidth obtained as a fit parameter from global analysis of the diagonal and antidiagonal cuts of the norm of the rephasing spectrum}  This linewidth is comparable to the inhomogeneous width of $\sim 6.5$\,meV, placing this system firmly in a dynamic disorder regime. Upon increasing exciton density, the homogeneous linewidth derived from the $\tau_p = 0$\,fs spectrum increases due to EID arising from many-body elastic scattering. As mentioned earlier in this section, by quantifying the contribution of EID to the homogeneous linewidth, we have have concluded that $X_A$ and $X_B$ demonstrate distinct exciton-exciton scattering rates, which we attributed to their peculiar phonon dressing~\cite{thouin2019phonon} leading to a specific dynamic Coulomb screening of their nonlinear coupling.

In Ref.~\citenum{thouin2019enhanced}, we considered the norm of the rephasing spectrum at $\tau_p = 0$\,fs; however upon close inspection of Fig.~\ref{fig:spectra1}, we notice that the real part of the lineshape in Fig.~\ref{fig:spectra1} displays dispersive shape, i.e.\ derivative shape about the peak energy, both for diagonal and off-diagonal resonances, in both the rephasing and non-rephasing spectrum. Note the sign-flip for the off-diagonal feature, which is consistent with its assignment to the excited state absorption to the biexcitonic state. Similarly, the imaginary part of the spectra (not shown in Fig.~\ref{fig:spectra1} but shown below in Fig.~\ref{fig:spectra2}) display an absorptive lineshape. The theoretical spectra shown in Fig.~\ref{fig:theoretical_lineshapes} suggest that such dispersive lineshapes are a consequence of many-body correlations. In the latter's absence, the lineshape should be purely absorptive. 
The spectra in Fig.~\ref{fig:spectra1} therefore reveal phase mixing due to many-body Coulomb correlations responsible for EID, as has been reported in semiconductor quantum wells~\cite{Li2006}.  In fact, we have demonstrated in Ref.~\citenum{thouin2019enhanced} that the EID dominates the non-linear response in the employed pump fluence range.  
These phenomena are reproduced by the 2D coherent spectra predicted by our stochastic theory, as shown in Fig.~\ref{fig:theoretical_lineshapes}.


The evolution of the rephasing lineshape shown in Fig.~\ref{fig:spectra1}(b) with population waiting time $\tau_p$ is displayed in Fig.~\ref{fig:spectra2}. The top row displays the real part of the spectrum at different values of $\tau_p$, the middle row the imaginary component, and the bottom row the norm (absolute value) of the complex spectrum. We observe that the phase scrambling phenomenon displayed in the $\tau_p=0$\,fs spectrum [Fig.~\ref{fig:spectra1}(b)] dissipates within $\tau_p \leq 240$\,fs: the real component of the spectrum evolves from an initially dispersive [Fig.~\ref{fig:spectra2}(a)] to absorptive [Fig.~\ref{fig:spectra2}(d)] lineshape, while that of the imaginary part evolves from absorptive [Fig.~\ref{fig:spectra2}(e)] to dispersive [Fig.~\ref{fig:spectra2}(h)] character. We note that although the evolution of the real and imaginary components of the complex lineshape is substantial over this ultrafast time window, the population decay of the diagonal features for $X_A$ and $X_B$ is weak, observed via the modest evolution of the total intensity in Fig.~\ref{fig:spectra2}(i)--(l). The decay of the $X_{A^{\prime}}$ diagonal peak and the biexciton cross peak appears more substantial. 

The marked evolution of the complex lineshape is also predicted by the stochastic theory as evidenced by the theoretical spectra presented in Fig.~\ref{fig:theory_time-resolved_spectra}. Recalling the arguments presented in Sec~\ref{sec:theo_pred}, such a dynamic is fundamentally driven by the exciton-exciton interactions that are time dependent due to the evolving background population. While the theory also predicts asymmetry in the lineshapes, non-negligible inhomegneous effects, which are not considered in theory yet present in the experimental spectra prevents the observation of clear asymmetry.

We also highlight the reduction in the total linewidth of the each diagonal exciton resonance in the absolute value of the response shown in Fig.~\ref{fig:spectra2}(i)--(l) with population time. 
Inspection of these spectra reveal dynamic narrowing of $X_A$ and $X_B$, primarily along the anti-diagonal spectral axis. It is more difficult to visually ascertain the linewidth evolution of $X_{A^{\prime}}$ and the biexciton cross peak given the non-negligible decay over this time period. The dynamic line narrowing in Fig.~\ref{fig:spectra2} has also been predicted by the theory, again due to the loss of dephasing pathway in the form of EID. 
This phenomenon reflects the spectral evolution predicted in 
\ref{fig:theory_width}, which highlights the dynamic line narrowing. 

To quantify the measured dynamic line narrowing, we display in Fig.~\ref{fig:width} the homogeneous linewidth as extracted in reference~\citenum{thouin2019enhanced} as a function of population time $\tau_p$. This is extracted by a global analysis of the diagonal and the antidiagonal lineshape as developed in references~\citenum{siemens2010resonance} and \citenum{bristow2011separating}: in the limit of similar homogeneous and inhomogeneous widths as is the case in this material~\cite{Thouin2018,Neutzner2018,thouin2019enhanced}, the diagonal lineshape follows a Voigt profile, while the antidiagonal spectrum is the product of a Gaussian and complementary error function, but both diagonal and antidiagonal widths depend on the dephasing parameter. By this analysis, Fig.~\ref{fig:width} shows that the linewidth of $X_A$ reduces most drastically, but that of $X_B$ also reduces over a typical time window, while $X_{A^{\prime}}$ displays no line narrowing. We note that in reference~\citenum{thouin2019enhanced}, we reported that $X_A$ has a stronger density dependence of EID than $X_B$, which is consistent with the observation derived from Figs.~\ref{fig:width}(a) and \ref{fig:width}(c). We have found $X_B$ to be more strongly displaced along phonon coordinates involving octahedral twist in the plane of the inorganic layer, and out of plane scissoring of the Pb---I---Pb apex~\cite{thouin2019phonon}. Interestingly, the homogeneous linewidth of $X_A$ displayed thermal broadening by a dominant phonon mode on the inorganic plane, while the thermal broadening mechanism for $X_B$ involved a phonon with motion involving the organic cation~\cite{Thouin2018}. The stronger exciton-phonon coupling implies that $X_B$ is more susceptible to dynamic screening than $X_A$, which is consistent with the data in Fig.~\ref{fig:width} and reference~\citenum{thouin2019enhanced}. Furthermore, the linewidth of $X_{A^{\prime}}$ displayed weaker, non-Boltzmann temperature dependence~\cite{thouin2019enhanced}. Finally, we point out that the asymptotic value of the homogeneous linewidth for $X_A$, $X_{A^{\prime}}$, and $X_B$ tends towards the low-exciton-density linewidths that we reported in reference~\citenum{thouin2019enhanced}. 

The linewidth of $X_{A^{\prime}}$ remains relatively constant over the probed population time. While this might initially suggest that this resonance is immune to EID effects, we note that the real part of the rephasing spectrum associated to this particular transition exhibits a dispersive lineshape at all population times, consistent with the inital lineshapes of $X_A$ and $X_B$. This indicates the clear presence of EID effects, as also confirmed by the density dependent linewidth previously published in Ref~\citenum{thouin2019enhanced}. The trend shown in Fig~\ref{fig:width}(b), on the other hand, suggests that the inter-exciton scattering does not evolve with the population time, at least within the probed time range. Inspection of the the lineshape, however, suggests that the dispersive lineshape of the real part is preserved at all population times, suggesting $X_{A^{\prime}}$ is subjected to EID over a much longer period of time than the other two resonances. Following the arguments developed by the theory in this paper, this implies the presence of a background exciton population that contributes to the scattering of $X_{A^{\prime}}$ and whose stochastic evolution is that of the background of the other two resonances. This reiterates our assignment of the multiple resonances within the spectral structure to excitonic states of distinct character and possibly specific origin~\cite{SrimathKandada2020}.

\subsection{Consequences for the exciton spectral structure}

The origin and nature of the spectral finestructure of the excitonic transition --- the presence of distinct resonances such as $X
_A$, $X_{A^{\prime}}$ and $X_B$ --- has been under discussion~\cite{mauck2019excitons}. Early works on 2D hybrid metal-halide perovskites suggested that the spectral structure is the outcome of degeneracy-lifting processes driven by spin-exchange interactions ubiquitous to lead-based semiconductors. Even though the spin-exchange energy was estimated to be of the same order magnitude as the energy spacing within the finestructure~\cite{Ema2006}, Kataoka et al.~\cite{kataoka1993magneto}, and more recently Urban et al.~\cite{urban2020revealing}, noted an indiscernible difference in the diamagnetic shift of each of the resonances with applied magnetic field. Alternatively, a more \textit{chemical} perspective was also suggested in which the structure was assigned to a vibronic progression within a single exciton state~\cite{urban2020revealing, Straus2018a}. This was particularly highlighted in \ce{(PEA)2PbI4}, where Urban et al.\cite{urban2020revealing} and Straus et al.\cite{straus2016direct} have identified a vibrational mode at about 40\,meV in the off-resonance Raman spectrum associated to the motion of the phenylethyammonium cation. We underline, however, that in order to unambiguously establish the vibronic nature, it is essential to measure the resonant Raman spectrum, which in fact has dominant contribution from the phonon modes of the inorganic lattice. We have reported that this type of measurement over a lower frequency range reveals distinct displacements along different phonon modes for $X_A$ and $X_B$~\cite{thouin2019phonon}.  Moreover, we highlight that similar, if not the same spectral structure, is observed even in other 2D hybrid metal-halide perovskite derivatives with other organic cations, further suggesting that the spectral structure is unlikely to be vibronic in nature.  

We have presented arguments in Ref.~\citenum{SrimathKandada2020} as to why we consider that such an interpretation of a vibronic progression for the spectral structure in Fig.~\ref{fig:spectra1}(a) does not explain a series of experimental observations, including distinct polaronic dressing of $X_A$ and $X_B$~\cite{thouin2019phonon}, distinct screening of EID for these excitons~\cite{thouin2019enhanced}, and distinct biexciton binding~\cite{Thouin2018}. We consider that the clearly peculiar behavior of $X_A$ and $X_B$ in Fig.~\ref{fig:width} adds to the body of work that establishes these excitons as a family of distinct but correlated excitons with strong polaronic character. 

\subsection{Summary of phenomenology and relationship to predictions from stochastic theory}

We summarize the phenomenology presented in ref.~\citenum{thouin2019enhanced} and in section~\ref{sec:2Dspec} pertaining to EID effects in the 2D coherent lineshape evolution in \ce{(PEA)2PbI4}.
\begin{itemize}
    \item The $\tau_p = 0$ rephasing absolute spectrum displays homogenous linewidth that depends on exciton density, with $X_A$, $X_{A^{\prime}}$, and $X_B$ displaying distinct density dependence~\cite{thouin2019enhanced}. Furthermore,  exciton-exciton scattering is activated by phonons on the lead iodide plane for $X_A$ but by motion of the organic cation for $X_B$, while $X_{A^{\prime}}$ displays relatively weaker thermal broadening~\cite{thouin2019enhanced}. We interpreted these distinct behaviors as indicative of peculiar screening by the lattice for each exciton polaron~\cite{thouin2019phonon}. 
    \item The model predicts a lineshape asymmetry, manifested both in the linear (ref.~\citenum{paper2}) and nonlinear (Fig.~\ref{fig:theoretical_lineshapes}) spectrum by a tail to higher energy that depends on the  background-exciton density $N_0$. This asymmetry is evident in the exciton-density dependence of the $\tau_p = 0$ rephasing lineshape measured experimentally (shown as Supplemental Material in ref.~\citenum{thouin2019enhanced}).
    \item Both the $\tau_p = 0$ rephasing and nonrephasing spectra display real and imaginary spectra with inverted lineshape: the real component displays dispersive lineshape while the imaginary one is absorptive (Figs.~\ref{fig:spectra1} and \ref{fig:spectra2}). This phenomenon is predicted by the stochastic model and arises from the background evolution from the $g_1$ term (Fig.~\ref{fig:theoretical_lineshapes}).
    \item This lineshape inversion dissipates over an ultrafast timescale in which the homogeneous linewidth decreases (Figs.~\ref{fig:spectra2} and \ref{fig:width}). The dynamics of the lineshape evolution are once again distinct for the different excitons. The linewidth evolution is predicted by the theory, and is ascribed to the time dependence of terms that arise due to EID effects (Figs.~\ref{fig:theory_time-resolved_spectra} and \ref{fig:theory_width}). In the situation of broadband excitation in which the initial distribution of exciton energies is broad, the model distinctly predicts dynamic line narrowing, but it also predicts dynamic line broadening in situations of narrow-band excitation (not shown in this manuscript). 
\end{itemize}
The microscopic model developed here captures the essential EID physics and opens new opportunities for novel, detailed understanding of many-body exciton physics and of how system-bath dynamics contribute to non-linear spectral behavior. We note that the theoretical model considered in the first part of the manuscript lacks some of the necessary ingredients to reproduce rigorously the experimental lineshapes, such as spin-exchange and polaronic effects. In spite of this limitation, we have qualitatively identified the physical origin of the observed nonlinear lineshape through the stochastic model presented in this article.

\section{Discussion}

We present here a joint theoretical and experimental 
study of excitation induced dephasing that connects the 
dynamics of an otherwise dark background density of states
to the evolution of the 2D coherent spectral lineshape. Such dynamics 
are input in our analytical model in the form of the spectral density
and corresponding stochastic equations of motion of the background population,
which dress the quantum operators for the ``bright'' degrees of freedom.
An important feature of our model and its connection to the experimental 
observation is that the complex and potentially intractable dynamics of the dark
variables can be reduced to a single stochastic variable and a few physical
parameters that can be directly related: exciton-exciton interactions and density of states. Coupled with It\^o calculus, this provides a powerful analytical tool
for interpreting dynamical features in 2D coherent spectra.  Furthermore, 
the approach can be extended to include additional interactions such as  polaronic binding and spin-orbit coupling. 
We reserve inclusion of these physics for future investigation. 

Our model predicts that the homogeneous linewidth evolves with population time purely due to the dynamics of many-body correlations. Dynamic line broadening effects in 2D coherent spectra are often interpreted in the context of spectral diffusion~\cite{roberts2006characterization,cho2008coherent}. This work demonstrates that in condensed matter systems, competing line narrowing processes due to many-body interactions can complicate such lineshape evolution. We consider that the fact that the stochastic model developed here and in Ref.~\citenum{paper2} predicts that the linewidth changes with time (whether it increases or decreases) is in itself a very important result with profound implications in condensed-matter chemical physics.

This stochastic theory as developed here is strictly valid for semiconductors, but it can be readily developed to include further physics appropriate for the description of Moir\'e excitons in 2D transition-metal dichalchogenide heterostructures~\cite{tang2020simulation}, and signatures of ground-state spin-orbit entanglement in the optical conductivity spectrum of quasi-one-dimensional Mott insulators~\cite{Controzzi2001,schlappa2012spin}, for example, These are the two optically accessible systems in which many-body interactions are dominant. Appropriate details of these physics can be, in principle, included in the spectral density in equation~\ref{eq:Nt}, and the consequences of these on nonstationary spectral behavior can be unravelled via our model.

\section{Perspective}

The spectral density of the environment plays a 
central role in many complex systems and governs the relaxation and 
decoherence of a quantum subsystem. Typically, we treat the environment 
as being quasi-stationary. The theoretical model presented in this article presents a means to represent the environment as nonstationary, here in the context of exciton-exciton scattering, leading to a rich evolution of the nonlinear coherent exciton lineshape. Our perspective is that there is ample scope to include richer physics in the spectral density, such as an explicit treatment of polaronic effects in materials such as the 2D metal-halide perovskites considered here~\cite{SrimathKandada2020}. We consider that this theoretical development presents opportunities to include microscopic understanding of many-body interactions that are dominant in condensed-matter systems on their quantum dynamics. 

The theoretical development presented here allows the exploration of the following open questions in chemical physics community: in multi-chromophoric systems such as light harvesting complexes,  
does sculpting of the spectral density determine sensitively (compared to experimental observables) the evolution of the optical exciton lineshape in 2D coherent excitation spectra? Does the additional microscopic detail contained in the  spectral density of the environment matter to capture the observed lineshape evolution? We believe that our theoretical framework can contribute towards this fundamental understanding.

\section*{Acknowledgements}
We are deeply thankful to Daniele Cortecchia and Annamaria Petrozza for providing the high quality samples for this work.  
The work at the University of Houston was funded in
part by the  National Science Foundation (CHE-1664971, DMR-1903785) and the Robert A. Welch Foundation (E-1337). The work at Georgia Tech was funded by the National Science Foundation (DMR-1904293). CS acknowledges support from the School of Chemistry and Biochemistry and the College of Science at Georgia Tech. 

\section*{Data Availability}

The data that support the findings of this study are available from the corresponding author upon reasonable request.

\appendix

\section{Experimental Methods}

\subsection{Sample Preparation}
The samples were provided by Dr.\ Daniele Cortecchia for the work presented in reference~\citenum{thouin2019enhanced}, which reports measurements taken concurrently with those reported in this article. Thin films of \ce{(PEA)2PbI4} (thickness of 40\,nm) were prepared on sapphire substrates (optical windows $25 \times 0.5\,$ mm, Crystran) by spin coating a 0.05\,M solution of the perovskite in  N,N-Dimethylformamide (DMF). A quantity of 12.5\,mg  of \ce{(PEA)I}  (Dyesol)  was  mixed  with  11.5\,mg \ce{PbI2} (TCI) and dissolved in 500\,$\mu$L of DMF (Sigma Aldrich, anhydrous, 99.8\%).  The solution was left to dissolve on a hotplate at $100^{\circ}$C for 1\,hr. After exposing the substrate to an oxygen plasma, the solution (kept at $100^{\circ}$C) was spin-coated on the sapphire window at 6000\,rpm for 30\,s, and the film was annealed on a hotplate at $100^{\circ}$C for 15 minutes.  The solution and the film were prepared in a glove-box under a \ce{N2}  atmosphere.

\subsection{Two-Dimensional Coherent Excitation Spectroscopy}
The pulse train (25\,fs, attenuated to a fluence $\sim 40$\,nJ/cm$^2$, centered at 530\,nm) was generated by a home-built single-pass non-collinear optical parametric amplifier pumped by the third harmonic of a Yb:KGW ultrafast laser system (Pharos Model PH1-20-0200-02-10, Light Conversion) with output pulse train at 1030\,nm and a repetition rate of 100\,kHz, an output power of 20\,W and a pulse duration of 220\,fs.  Two-dimensional spectroscopic measurements were performed using a home-built, pulse-shaper-based multidimensional spectrometer that passively stabilizes the relative phase of each pulses~\cite{turner2011invited}. Our implementation is described in detail in ref.~\citenum{Thouin2018}, albeit with a different ultrafast laser source with much lower repetition rate.  Each beams was independently compressed using chirp-scan~\cite{loriot2013self} to a pulse duration of 25\,fs FWHM, and was characterized using cross-correlated second harmonic frequency resolved optical gating~\cite{trebino2012frequency} (SH-XFROG) in a 10\,$\mu$m-thick BBO crystal placed at the sample position.  For details of pulse characterization, including a typical SH-XFROG trace, we refer the reader to reference~\citenum{thouin2019enhanced} since the data presented in this manuscript was taken simultaneously as the data presented in that article.  The sample was kept at 5\,K using a vibration-free cold-finger closed-cycle cryostat (Montana Instruments).

\section{Correlation Functions}

Here we provide the third-order correlation functions have been used in this work. The correlation functions are defined in the same way as in the reference~\citenum{Mukamel1995}. The subscripts ``$a$'' and ``$b$'' denote single and double excitation manifolds, respectively. 

\begin{widetext}
The correlation functions with rephasing phase matching are
\begin{align}
R_{2a}(t_3,t_2,t_1) 
=& R_{3a}(t_3,t_2,t_1) \\ \nonumber 
=&\exp\left[-\frac{i}{\hbar}(\omega_0 + n_0 \frac{\gamma_1}{2})(t_3 - t_1)\right] \exp\left[-\frac{i}{\hbar}\frac{\gamma_1 N_0}{\gamma}(e^{-\gamma t_1}+e^{-\gamma(t_1+t_2)}-e^{-\gamma(t_1+t_2+t_3)}-1)\right]\\ \nonumber
&\times \exp\left[-\frac{\gamma_1^2 \sigma^2}{4\hbar^2\gamma^3}\left(-e^{-2\gamma t_1} +4e^{-\gamma t_1} -2e^{-\gamma t_2} -e^{-2\gamma(t_1+t_2)} +4e^{-\gamma(t_1+t_2)} -2e^{-\gamma(2t_1+t_2)} + 2e^{-\gamma t_3} +2e^{-\gamma(t_2+t_3)}\right.\right.\\ \nonumber
&\qquad \qquad \qquad \quad \left.\left. -e^{-2\gamma(t_1+t_2+t_3)} -4e^{-\gamma(t_1+t_2+t_3)} +2e^{-\gamma(2t_1+t_2+t_3)} +2e^{-\gamma(2t_1+2t_2+t_3)} +2\gamma(t_1+t_3) -5\right)\right.\\ \nonumber
&\qquad \quad \left.- \frac{\gamma_1^2 \sigma_{N_o}^2}{2\hbar^2\gamma^2} \left(e^{-\gamma t_1} +e^{-\gamma(t_1+t_2)} -e^{-\gamma(t_1+t_2+t_3)}-1\right)^2 \right],
\end{align}
and 
\begin{align}
R_{1b}^*(t_3,t_2,t_1)=2\exp\left[-\frac{i}{2\hbar}\gamma_1 t_3\right]R_{2a}(t_3,t_2,t_1).
\end{align}

The correlation functions with non-rephasing phase matching are
\begin{align}
R_{1a}(t_3,t_2,t_1)= & R_{4a}(t_3,t_2,t_1)\\ \nonumber
=&\exp\left[-\frac{i}{\hbar}(\omega_0 + n_0 \frac{\gamma_1}{2})(t_3 + t_1)\right] \exp\left[-\frac{i}{\hbar}\frac{\gamma_1 N_0}{\gamma}(-e^{-\gamma t_1}+e^{-\gamma(t_1+t_2)}-e^{-\gamma(t_1+t_2+t_3)}+1)\right]\\ \nonumber
&\times \exp\left[-\frac{\gamma_1^2 \sigma^2}{4\hbar^2\gamma^3}\left(-e^{-2\gamma t_1} +4e^{-\gamma t_1} +2e^{-\gamma t_2} -e^{-2\gamma(t_1+t_2)} -4e^{-\gamma(t_1+t_2)} +2e^{-\gamma(2t_1+t_2)} + 2e^{-\gamma t_3} -2e^{-\gamma(t_2+t_3)}\right.\right.\\ \nonumber
&\qquad \qquad \qquad \quad \left.\left. -e^{-2\gamma(t_1+t_2+t_3)} +4e^{-\gamma(t_1+t_2+t_3)} -2e^{-\gamma(2t_1+t_2+t_3)} +2e^{-\gamma(2t_1+2t_2+t_3)} +2\gamma(t_1+t_3) -5\right)\right.\\ \nonumber
&\qquad \quad \left.- \frac{\gamma_1^2 \sigma_{N_o}^2}{2\hbar^2\gamma^2} \left(-e^{-\gamma t_1} +e^{-\gamma(t_1+t_2)} -e^{-\gamma(t_1+t_2+t_3)}+1\right)^2 \right],
\end{align}
and
\begin{align}
R_{2b}^*(t_3,t_2,t_1)=2\exp\left[-\frac{i}{2\hbar}\gamma_1 t_3\right]R_{1a}(t_3,t_2,t_1).
\end{align}
\end{widetext}

\section*{References}

\begin{thebibliography}{66}%
	\makeatletter
	\providecommand \@ifxundefined [1]{%
		\@ifx{#1\undefined}
	}%
	\providecommand \@ifnum [1]{%
		\ifnum #1\expandafter \@firstoftwo
		\else \expandafter \@secondoftwo
		\fi
	}%
	\providecommand \@ifx [1]{%
		\ifx #1\expandafter \@firstoftwo
		\else \expandafter \@secondoftwo
		\fi
	}%
	\providecommand \natexlab [1]{#1}%
	\providecommand \enquote  [1]{``#1''}%
	\providecommand \bibnamefont  [1]{#1}%
	\providecommand \bibfnamefont [1]{#1}%
	\providecommand \citenamefont [1]{#1}%
	\providecommand \href@noop [0]{\@secondoftwo}%
	\providecommand \href [0]{\begingroup \@sanitize@url \@href}%
	\providecommand \@href[1]{\@@startlink{#1}\@@href}%
	\providecommand \@@href[1]{\endgroup#1\@@endlink}%
	\providecommand \@sanitize@url [0]{\catcode `\\12\catcode `\$12\catcode
		`\&12\catcode `\#12\catcode `\^12\catcode `\_12\catcode `\%12\relax}%
	\providecommand \@@startlink[1]{}%
	\providecommand \@@endlink[0]{}%
	\providecommand \url  [0]{\begingroup\@sanitize@url \@url }%
	\providecommand \@url [1]{\endgroup\@href {#1}{\urlprefix }}%
	\providecommand \urlprefix  [0]{URL }%
	\providecommand \Eprint [0]{\href }%
	\providecommand \doibase [0]{https://doi.org/}%
	\providecommand \selectlanguage [0]{\@gobble}%
	\providecommand \bibinfo  [0]{\@secondoftwo}%
	\providecommand \bibfield  [0]{\@secondoftwo}%
	\providecommand \translation [1]{[#1]}%
	\providecommand \BibitemOpen [0]{}%
	\providecommand \bibitemStop [0]{}%
	\providecommand \bibitemNoStop [0]{.\EOS\space}%
	\providecommand \EOS [0]{\spacefactor3000\relax}%
	\providecommand \BibitemShut  [1]{\csname bibitem#1\endcsname}%
	\let\auto@bib@innerbib\@empty
	\bibitem [{\citenamefont {Mysyrowicz}\ \emph {et~al.}(1968)\citenamefont
		{Mysyrowicz}, \citenamefont {Grun}, \citenamefont {Levy}, \citenamefont
		{Bivas},\ and\ \citenamefont {Nikitine}}]{mysyrowicz1968excitonic}%
	\BibitemOpen
	\bibfield  {author} {\bibinfo {author} {\bibfnamefont {A.}~\bibnamefont
			{Mysyrowicz}}, \bibinfo {author} {\bibfnamefont {J.}~\bibnamefont {Grun}},
		\bibinfo {author} {\bibfnamefont {R.}~\bibnamefont {Levy}}, \bibinfo {author}
		{\bibfnamefont {A.}~\bibnamefont {Bivas}},\ and\ \bibinfo {author}
		{\bibfnamefont {S.}~\bibnamefont {Nikitine}},\ }\bibfield  {title} {\enquote
		{\bibinfo {title} {Excitonic molecule in {CuC1}},}\ }\href@noop {} {\bibfield
		{journal} {\bibinfo  {journal} {Phys. Lett. A}\ }\textbf {\bibinfo {volume}
			{26}},\ \bibinfo {pages} {615--616} (\bibinfo {year} {1968})}\BibitemShut
	{NoStop}%
	\bibitem [{\citenamefont {Magde}\ and\ \citenamefont
		{Mahr}(1970)}]{magde1970exciton}%
	\BibitemOpen
	\bibfield  {author} {\bibinfo {author} {\bibfnamefont {D.}~\bibnamefont
			{Magde}}\ and\ \bibinfo {author} {\bibfnamefont {H.}~\bibnamefont {Mahr}},\
	}\bibfield  {title} {\enquote {\bibinfo {title} {Exciton-exciton interaction
				in {CdS, CdSe, and ZnO}},}\ }\href@noop {} {\bibfield  {journal} {\bibinfo
			{journal} {Phys. Rev. Lett.}\ }\textbf {\bibinfo {volume} {24}},\ \bibinfo
		{pages} {890} (\bibinfo {year} {1970})}\BibitemShut {NoStop}%
	\bibitem [{\citenamefont {Grun}\ \emph {et~al.}(1970)\citenamefont {Grun},
		\citenamefont {Nikitine}, \citenamefont {Bivas},\ and\ \citenamefont
		{Levy}}]{grun1970luminescence}%
	\BibitemOpen
	\bibfield  {author} {\bibinfo {author} {\bibfnamefont {J.}~\bibnamefont
			{Grun}}, \bibinfo {author} {\bibfnamefont {S.}~\bibnamefont {Nikitine}},
		\bibinfo {author} {\bibfnamefont {A.}~\bibnamefont {Bivas}},\ and\ \bibinfo
		{author} {\bibfnamefont {R.}~\bibnamefont {Levy}},\ }\bibfield  {title}
	{\enquote {\bibinfo {title} {Luminescence of copper halides excited by a high
				power laser},}\ }\href@noop {} {\bibfield  {journal} {\bibinfo  {journal} {J.
				Lumin.}\ }\textbf {\bibinfo {volume} {1}},\ \bibinfo {pages} {241--253}
		(\bibinfo {year} {1970})}\BibitemShut {NoStop}%
	\bibitem [{\citenamefont {Miller}\ \emph {et~al.}(1982)\citenamefont {Miller},
		\citenamefont {Kleinman}, \citenamefont {Gossard},\ and\ \citenamefont
		{Munteanu}}]{miller1982biexcitons}%
	\BibitemOpen
	\bibfield  {author} {\bibinfo {author} {\bibfnamefont {R.}~\bibnamefont
			{Miller}}, \bibinfo {author} {\bibfnamefont {D.}~\bibnamefont {Kleinman}},
		\bibinfo {author} {\bibfnamefont {A.}~\bibnamefont {Gossard}},\ and\ \bibinfo
		{author} {\bibfnamefont {O.}~\bibnamefont {Munteanu}},\ }\bibfield  {title}
	{\enquote {\bibinfo {title} {Biexcitons in {GaAs} quantum wells},}\
	}\href@noop {} {\bibfield  {journal} {\bibinfo  {journal} {Phys. Rev. B}\
		}\textbf {\bibinfo {volume} {25}},\ \bibinfo {pages} {6545} (\bibinfo {year}
		{1982})}\BibitemShut {NoStop}%
	\bibitem [{\citenamefont {Kleinman}(1983)}]{kleinman1983binding}%
	\BibitemOpen
	\bibfield  {author} {\bibinfo {author} {\bibfnamefont {D.}~\bibnamefont
			{Kleinman}},\ }\bibfield  {title} {\enquote {\bibinfo {title} {Binding energy
				of biexcitons and bound excitons in quantum wells},}\ }\href@noop {}
	{\bibfield  {journal} {\bibinfo  {journal} {Phys. Rev. B}\ }\textbf {\bibinfo
			{volume} {28}},\ \bibinfo {pages} {871} (\bibinfo {year} {1983})}\BibitemShut
	{NoStop}%
	\bibitem [{\citenamefont {Hu}\ \emph {et~al.}(1990)\citenamefont {Hu},
		\citenamefont {Koch}, \citenamefont {Lindberg}, \citenamefont
		{Peyghambarian}, \citenamefont {Pollock},\ and\ \citenamefont
		{Abraham}}]{hu1990biexcitons}%
	\BibitemOpen
	\bibfield  {author} {\bibinfo {author} {\bibfnamefont {Y.}~\bibnamefont
			{Hu}}, \bibinfo {author} {\bibfnamefont {S.~W.}\ \bibnamefont {Koch}},
		\bibinfo {author} {\bibfnamefont {M.}~\bibnamefont {Lindberg}}, \bibinfo
		{author} {\bibfnamefont {N.}~\bibnamefont {Peyghambarian}}, \bibinfo {author}
		{\bibfnamefont {E.}~\bibnamefont {Pollock}},\ and\ \bibinfo {author}
		{\bibfnamefont {F.~F.}\ \bibnamefont {Abraham}},\ }\bibfield  {title}
	{\enquote {\bibinfo {title} {Biexcitons in semiconductor quantum dots},}\
	}\href@noop {} {\bibfield  {journal} {\bibinfo  {journal} {Phys. Rev. Lett.}\
		}\textbf {\bibinfo {volume} {64}},\ \bibinfo {pages} {1805} (\bibinfo {year}
		{1990})}\BibitemShut {NoStop}%
	\bibitem [{\citenamefont {Brunner}\ \emph {et~al.}(1994)\citenamefont
		{Brunner}, \citenamefont {Abstreiter}, \citenamefont {B{\"o}hm},
		\citenamefont {Tr{\"a}nkle},\ and\ \citenamefont
		{Weimann}}]{brunner1994sharp}%
	\BibitemOpen
	\bibfield  {author} {\bibinfo {author} {\bibfnamefont {K.}~\bibnamefont
			{Brunner}}, \bibinfo {author} {\bibfnamefont {G.}~\bibnamefont {Abstreiter}},
		\bibinfo {author} {\bibfnamefont {G.}~\bibnamefont {B{\"o}hm}}, \bibinfo
		{author} {\bibfnamefont {G.}~\bibnamefont {Tr{\"a}nkle}},\ and\ \bibinfo
		{author} {\bibfnamefont {G.}~\bibnamefont {Weimann}},\ }\bibfield  {title}
	{\enquote {\bibinfo {title} {Sharp-line photoluminescence and two-photon
				absorption of zero-dimensional biexcitons in a {GaAs/AlGaAs} structure},}\
	}\href@noop {} {\bibfield  {journal} {\bibinfo  {journal} {Phys. Rev. Lett.}\
		}\textbf {\bibinfo {volume} {73}},\ \bibinfo {pages} {1138} (\bibinfo {year}
		{1994})}\BibitemShut {NoStop}%
	\bibitem [{\citenamefont {Albrecht}\ \emph {et~al.}(1996)\citenamefont
		{Albrecht}, \citenamefont {Bott}, \citenamefont {Meier}, \citenamefont
		{Schulze}, \citenamefont {Koch}, \citenamefont {Cundiff}, \citenamefont
		{Feldmann}, \citenamefont {Stolz}, \citenamefont {Thomas}, \citenamefont
		{Koch} \emph {et~al.}}]{albrecht1996disorder}%
	\BibitemOpen
	\bibfield  {author} {\bibinfo {author} {\bibfnamefont {T.}~\bibnamefont
			{Albrecht}}, \bibinfo {author} {\bibfnamefont {K.}~\bibnamefont {Bott}},
		\bibinfo {author} {\bibfnamefont {T.}~\bibnamefont {Meier}}, \bibinfo
		{author} {\bibfnamefont {A.}~\bibnamefont {Schulze}}, \bibinfo {author}
		{\bibfnamefont {M.}~\bibnamefont {Koch}}, \bibinfo {author} {\bibfnamefont
			{S.}~\bibnamefont {Cundiff}}, \bibinfo {author} {\bibfnamefont
			{J.}~\bibnamefont {Feldmann}}, \bibinfo {author} {\bibfnamefont
			{W.}~\bibnamefont {Stolz}}, \bibinfo {author} {\bibfnamefont
			{P.}~\bibnamefont {Thomas}}, \bibinfo {author} {\bibfnamefont {S.~W.}\
			\bibnamefont {Koch}}, \emph {et~al.},\ }\bibfield  {title} {\enquote
		{\bibinfo {title} {Disorder mediated biexcitonic beats in semiconductor
				quantum wells},}\ }\href@noop {} {\bibfield  {journal} {\bibinfo  {journal}
			{Phys. Rev. B}\ }\textbf {\bibinfo {volume} {54}},\ \bibinfo {pages} {4436}
		(\bibinfo {year} {1996})}\BibitemShut {NoStop}%
	\bibitem [{\citenamefont {Stone}\ \emph {et~al.}(2009)\citenamefont {Stone},
		\citenamefont {Gundogdu}, \citenamefont {Turner}, \citenamefont {Li},
		\citenamefont {Cundiff},\ and\ \citenamefont {Nelson}}]{stone2009two}%
	\BibitemOpen
	\bibfield  {author} {\bibinfo {author} {\bibfnamefont {K.~W.}\ \bibnamefont
			{Stone}}, \bibinfo {author} {\bibfnamefont {K.}~\bibnamefont {Gundogdu}},
		\bibinfo {author} {\bibfnamefont {D.~B.}\ \bibnamefont {Turner}}, \bibinfo
		{author} {\bibfnamefont {X.}~\bibnamefont {Li}}, \bibinfo {author}
		{\bibfnamefont {S.~T.}\ \bibnamefont {Cundiff}},\ and\ \bibinfo {author}
		{\bibfnamefont {K.~A.}\ \bibnamefont {Nelson}},\ }\bibfield  {title}
	{\enquote {\bibinfo {title} {Two-quantum 2d ft electronic spectroscopy of
				biexcitons in gaas quantum wells},}\ }\href@noop {} {\bibfield  {journal}
		{\bibinfo  {journal} {Science}\ }\textbf {\bibinfo {volume} {324}},\ \bibinfo
		{pages} {1169--1173} (\bibinfo {year} {2009})}\BibitemShut {NoStop}%
	\bibitem [{\citenamefont {Karaiskaj}\ \emph {et~al.}(2010)\citenamefont
		{Karaiskaj}, \citenamefont {Bristow}, \citenamefont {Yang}, \citenamefont
		{Dai}, \citenamefont {Mirin}, \citenamefont {Mukamel},\ and\ \citenamefont
		{Cundiff}}]{karaiskaj2010two}%
	\BibitemOpen
	\bibfield  {author} {\bibinfo {author} {\bibfnamefont {D.}~\bibnamefont
			{Karaiskaj}}, \bibinfo {author} {\bibfnamefont {A.~D.}\ \bibnamefont
			{Bristow}}, \bibinfo {author} {\bibfnamefont {L.}~\bibnamefont {Yang}},
		\bibinfo {author} {\bibfnamefont {X.}~\bibnamefont {Dai}}, \bibinfo {author}
		{\bibfnamefont {R.~P.}\ \bibnamefont {Mirin}}, \bibinfo {author}
		{\bibfnamefont {S.}~\bibnamefont {Mukamel}},\ and\ \bibinfo {author}
		{\bibfnamefont {S.~T.}\ \bibnamefont {Cundiff}},\ }\bibfield  {title}
	{\enquote {\bibinfo {title} {Two-quantum many-body coherences in
				two-dimensional fourier-transform spectra of exciton resonances in
				semiconductor quantum wells},}\ }\href@noop {} {\bibfield  {journal}
		{\bibinfo  {journal} {Phys. Rev. Lett.}\ }\textbf {\bibinfo {volume} {104}},\
		\bibinfo {pages} {117401} (\bibinfo {year} {2010})}\BibitemShut {NoStop}%
	\bibitem [{\citenamefont {Turner}\ and\ \citenamefont
		{Nelson}(2010)}]{turner2010coherent}%
	\BibitemOpen
	\bibfield  {author} {\bibinfo {author} {\bibfnamefont {D.~B.}\ \bibnamefont
			{Turner}}\ and\ \bibinfo {author} {\bibfnamefont {K.~A.}\ \bibnamefont
			{Nelson}},\ }\bibfield  {title} {\enquote {\bibinfo {title} {Coherent
				measurements of high-order electronic correlations in quantum wells},}\
	}\href@noop {} {\bibfield  {journal} {\bibinfo  {journal} {Nature}\ }\textbf
		{\bibinfo {volume} {466}},\ \bibinfo {pages} {1089--1092} (\bibinfo {year}
		{2010})}\BibitemShut {NoStop}%
	\bibitem [{\citenamefont {Schultheis}\ \emph {et~al.}(1986)\citenamefont
		{Schultheis}, \citenamefont {Kuhl}, \citenamefont {Honold},\ and\
		\citenamefont {Tu}}]{Schultheis1986}%
	\BibitemOpen
	\bibfield  {author} {\bibinfo {author} {\bibfnamefont {L.}~\bibnamefont
			{Schultheis}}, \bibinfo {author} {\bibfnamefont {J.}~\bibnamefont {Kuhl}},
		\bibinfo {author} {\bibfnamefont {A.}~\bibnamefont {Honold}},\ and\ \bibinfo
		{author} {\bibfnamefont {C.~W.}\ \bibnamefont {Tu}},\ }\bibfield  {title}
	{\enquote {\bibinfo {title} {Ultrafast phase relaxation of excitons via
				exciton-exciton and exciton-electron collisions},}\ }\href
	{https://doi.org/10.1103/PhysRevLett.57.1635} {\bibfield  {journal} {\bibinfo
			{journal} {Phys. Rev. Lett.}\ }\textbf {\bibinfo {volume} {57}},\ \bibinfo
		{pages} {1635--1638} (\bibinfo {year} {1986})}\BibitemShut {NoStop}%
	\bibitem [{\citenamefont {Honold}\ \emph {et~al.}(1989)\citenamefont {Honold},
		\citenamefont {Schultheis}, \citenamefont {Kuhl},\ and\ \citenamefont
		{Tu}}]{Honold1989}%
	\BibitemOpen
	\bibfield  {author} {\bibinfo {author} {\bibfnamefont {A.}~\bibnamefont
			{Honold}}, \bibinfo {author} {\bibfnamefont {L.}~\bibnamefont {Schultheis}},
		\bibinfo {author} {\bibfnamefont {J.}~\bibnamefont {Kuhl}},\ and\ \bibinfo
		{author} {\bibfnamefont {C.~W.}\ \bibnamefont {Tu}},\ }\bibfield  {title}
	{\enquote {\bibinfo {title} {Collision broadening of two-dimensional excitons
				in a {GaAs} single quantum well},}\ }\href
	{https://doi.org/10.1103/PhysRevB.40.6442} {\bibfield  {journal} {\bibinfo
			{journal} {Phys. Rev. B}\ }\textbf {\bibinfo {volume} {40}},\ \bibinfo
		{pages} {6442--6445} (\bibinfo {year} {1989})}\BibitemShut {NoStop}%
	\bibitem [{\citenamefont {Wang}\ \emph {et~al.}(1993)\citenamefont {Wang},
		\citenamefont {Ferrio}, \citenamefont {Steel}, \citenamefont {Hu},
		\citenamefont {Binder},\ and\ \citenamefont {Koch}}]{Wang1993}%
	\BibitemOpen
	\bibfield  {author} {\bibinfo {author} {\bibfnamefont {H.}~\bibnamefont
			{Wang}}, \bibinfo {author} {\bibfnamefont {K.}~\bibnamefont {Ferrio}},
		\bibinfo {author} {\bibfnamefont {D.~G.}\ \bibnamefont {Steel}}, \bibinfo
		{author} {\bibfnamefont {Y.~Z.}\ \bibnamefont {Hu}}, \bibinfo {author}
		{\bibfnamefont {R.}~\bibnamefont {Binder}},\ and\ \bibinfo {author}
		{\bibfnamefont {S.~W.}\ \bibnamefont {Koch}},\ }\bibfield  {title} {\enquote
		{\bibinfo {title} {Transient nonlinear optical response from excitation
				induced dephasing in {GaAs}},}\ }\href
	{https://doi.org/10.1103/PhysRevLett.71.1261} {\bibfield  {journal} {\bibinfo
			{journal} {Phys. Rev. Lett.}\ }\textbf {\bibinfo {volume} {71}},\ \bibinfo
		{pages} {1261--1264} (\bibinfo {year} {1993})}\BibitemShut {NoStop}%
	\bibitem [{\citenamefont {Wang}\ \emph {et~al.}(1994)\citenamefont {Wang},
		\citenamefont {Ferrio}, \citenamefont {Steel}, \citenamefont {Berman},
		\citenamefont {Hu}, \citenamefont {Binder},\ and\ \citenamefont
		{Koch}}]{Wang1994}%
	\BibitemOpen
	\bibfield  {author} {\bibinfo {author} {\bibfnamefont {H.}~\bibnamefont
			{Wang}}, \bibinfo {author} {\bibfnamefont {K.~B.}\ \bibnamefont {Ferrio}},
		\bibinfo {author} {\bibfnamefont {D.~G.}\ \bibnamefont {Steel}}, \bibinfo
		{author} {\bibfnamefont {P.~R.}\ \bibnamefont {Berman}}, \bibinfo {author}
		{\bibfnamefont {Y.~Z.}\ \bibnamefont {Hu}}, \bibinfo {author} {\bibfnamefont
			{R.}~\bibnamefont {Binder}},\ and\ \bibinfo {author} {\bibfnamefont {S.~W.}\
			\bibnamefont {Koch}},\ }\bibfield  {title} {\enquote {\bibinfo {title}
			{Transient four-wave-mixing line shapes: Effects of excitation-induced
				dephasing},}\ }\href {https://doi.org/10.1103/PhysRevA.49.R1551} {\bibfield
		{journal} {\bibinfo  {journal} {Phys. Rev. A}\ }\textbf {\bibinfo {volume}
			{49}},\ \bibinfo {pages} {R1551--R1554} (\bibinfo {year} {1994})}\BibitemShut
	{NoStop}%
	\bibitem [{\citenamefont {Hu}\ \emph {et~al.}(1994)\citenamefont {Hu},
		\citenamefont {Binder}, \citenamefont {Koch}, \citenamefont {Cundiff},
		\citenamefont {Wang},\ and\ \citenamefont {Steel}}]{Hu1994}%
	\BibitemOpen
	\bibfield  {author} {\bibinfo {author} {\bibfnamefont {Y.~Z.}\ \bibnamefont
			{Hu}}, \bibinfo {author} {\bibfnamefont {R.}~\bibnamefont {Binder}}, \bibinfo
		{author} {\bibfnamefont {S.~W.}\ \bibnamefont {Koch}}, \bibinfo {author}
		{\bibfnamefont {S.~T.}\ \bibnamefont {Cundiff}}, \bibinfo {author}
		{\bibfnamefont {H.}~\bibnamefont {Wang}},\ and\ \bibinfo {author}
		{\bibfnamefont {D.~G.}\ \bibnamefont {Steel}},\ }\bibfield  {title} {\enquote
		{\bibinfo {title} {Excitation and polarization effects in semiconductor
				four-wave-mixing spectroscopy},}\ }\href
	{https://doi.org/10.1103/PhysRevB.49.14382} {\bibfield  {journal} {\bibinfo
			{journal} {Phys. Rev. B}\ }\textbf {\bibinfo {volume} {49}},\ \bibinfo
		{pages} {14382--14386} (\bibinfo {year} {1994})}\BibitemShut {NoStop}%
	\bibitem [{\citenamefont {Rappen}\ \emph {et~al.}(1994)\citenamefont {Rappen},
		\citenamefont {Peter}, \citenamefont {Wegener},\ and\ \citenamefont
		{Sch\"afer}}]{Rappen1994}%
	\BibitemOpen
	\bibfield  {author} {\bibinfo {author} {\bibfnamefont {T.}~\bibnamefont
			{Rappen}}, \bibinfo {author} {\bibfnamefont {U.-G.}\ \bibnamefont {Peter}},
		\bibinfo {author} {\bibfnamefont {M.}~\bibnamefont {Wegener}},\ and\ \bibinfo
		{author} {\bibfnamefont {W.}~\bibnamefont {Sch\"afer}},\ }\bibfield  {title}
	{\enquote {\bibinfo {title} {Polarization dependence of dephasing processes:
				A probe for many-body effects},}\ }\href
	{https://doi.org/10.1103/PhysRevB.49.10774} {\bibfield  {journal} {\bibinfo
			{journal} {Phys. Rev. B}\ }\textbf {\bibinfo {volume} {49}},\ \bibinfo
		{pages} {10774--10777} (\bibinfo {year} {1994})}\BibitemShut {NoStop}%
	\bibitem [{\citenamefont {Wagner}\ \emph {et~al.}(1997)\citenamefont {Wagner},
		\citenamefont {Sch\"atz}, \citenamefont {Maier}, \citenamefont {Langbein},\
		and\ \citenamefont {Hvam}}]{Wagner1997}%
	\BibitemOpen
	\bibfield  {author} {\bibinfo {author} {\bibfnamefont {H.~P.}\ \bibnamefont
			{Wagner}}, \bibinfo {author} {\bibfnamefont {A.}~\bibnamefont {Sch\"atz}},
		\bibinfo {author} {\bibfnamefont {R.}~\bibnamefont {Maier}}, \bibinfo
		{author} {\bibfnamefont {W.}~\bibnamefont {Langbein}},\ and\ \bibinfo
		{author} {\bibfnamefont {J.~M.}\ \bibnamefont {Hvam}},\ }\bibfield  {title}
	{\enquote {\bibinfo {title} {Coherent optical nonlinearities and phase
				relaxation of quasi-three-dimensional and quasi-two-dimensional excitons in
				{ZnS$_{x}$Se$_{1-x}$/ZnSe} structures},}\ }\href
	{https://doi.org/10.1103/PhysRevB.56.12581} {\bibfield  {journal} {\bibinfo
			{journal} {Phys. Rev. B}\ }\textbf {\bibinfo {volume} {56}},\ \bibinfo
		{pages} {12581--12588} (\bibinfo {year} {1997})}\BibitemShut {NoStop}%
	\bibitem [{\citenamefont {Wagner}\ \emph {et~al.}(1999)\citenamefont {Wagner},
		\citenamefont {Sch\"atz}, \citenamefont {Langbein}, \citenamefont {Hvam},\
		and\ \citenamefont {Smirl}}]{Wagner1999}%
	\BibitemOpen
	\bibfield  {author} {\bibinfo {author} {\bibfnamefont {H.~P.}\ \bibnamefont
			{Wagner}}, \bibinfo {author} {\bibfnamefont {A.}~\bibnamefont {Sch\"atz}},
		\bibinfo {author} {\bibfnamefont {W.}~\bibnamefont {Langbein}}, \bibinfo
		{author} {\bibfnamefont {J.~M.}\ \bibnamefont {Hvam}},\ and\ \bibinfo
		{author} {\bibfnamefont {A.~L.}\ \bibnamefont {Smirl}},\ }\bibfield  {title}
	{\enquote {\bibinfo {title} {Interaction-induced effects in the nonlinear
				coherent response of quantum-well excitons},}\ }\href
	{https://doi.org/10.1103/PhysRevB.60.4454} {\bibfield  {journal} {\bibinfo
			{journal} {Phys. Rev. B}\ }\textbf {\bibinfo {volume} {60}},\ \bibinfo
		{pages} {4454--4457} (\bibinfo {year} {1999})}\BibitemShut {NoStop}%
	\bibitem [{\citenamefont {Shacklette}\ and\ \citenamefont
		{Cundiff}(2002)}]{shacklette2002role}%
	\BibitemOpen
	\bibfield  {author} {\bibinfo {author} {\bibfnamefont {J.~M.}\ \bibnamefont
			{Shacklette}}\ and\ \bibinfo {author} {\bibfnamefont {S.~T.}\ \bibnamefont
			{Cundiff}},\ }\bibfield  {title} {\enquote {\bibinfo {title} {Role of
				excitation-induced shift in the coherent optical response of
				semiconductors},}\ }\href@noop {} {\bibfield  {journal} {\bibinfo  {journal}
			{Phys. Rev. B}\ }\textbf {\bibinfo {volume} {66}},\ \bibinfo {pages} {045309}
		(\bibinfo {year} {2002})}\BibitemShut {NoStop}%
	\bibitem [{\citenamefont {Shacklette}\ and\ \citenamefont
		{Cundiff}(2003)}]{shacklette2003nonperturbative}%
	\BibitemOpen
	\bibfield  {author} {\bibinfo {author} {\bibfnamefont {J.}~\bibnamefont
			{Shacklette}}\ and\ \bibinfo {author} {\bibfnamefont {S.~T.}\ \bibnamefont
			{Cundiff}},\ }\bibfield  {title} {\enquote {\bibinfo {title} {Nonperturbative
				transient four-wave-mixing line shapes due to excitation-induced shift and
				excitation-induced dephasing},}\ }\href@noop {} {\bibfield  {journal}
		{\bibinfo  {journal} {J. Opt. Soc. Am. B}\ }\textbf {\bibinfo {volume}
			{20}},\ \bibinfo {pages} {764--769} (\bibinfo {year} {2003})}\BibitemShut
	{NoStop}%
	\bibitem [{\citenamefont {Li}\ \emph {et~al.}(2006)\citenamefont {Li},
		\citenamefont {Zhang}, \citenamefont {Borca},\ and\ \citenamefont
		{Cundiff}}]{Li2006}%
	\BibitemOpen
	\bibfield  {author} {\bibinfo {author} {\bibfnamefont {X.}~\bibnamefont
			{Li}}, \bibinfo {author} {\bibfnamefont {T.}~\bibnamefont {Zhang}}, \bibinfo
		{author} {\bibfnamefont {C.~N.}\ \bibnamefont {Borca}},\ and\ \bibinfo
		{author} {\bibfnamefont {S.~T.}\ \bibnamefont {Cundiff}},\ }\bibfield
	{title} {\enquote {\bibinfo {title} {Many-body interactions in semiconductors
				probed by optical two-dimensional fourier transform spectroscopy},}\ }\href
	{https://doi.org/10.1103/PhysRevLett.96.057406} {\bibfield  {journal}
		{\bibinfo  {journal} {Phys. Rev. Lett.}\ }\textbf {\bibinfo {volume} {96}},\
		\bibinfo {pages} {057406} (\bibinfo {year} {2006})}\BibitemShut {NoStop}%
	\bibitem [{\citenamefont {Moody}\ \emph {et~al.}(2011)\citenamefont {Moody},
		\citenamefont {Siemens}, \citenamefont {Bristow}, \citenamefont {Dai},
		\citenamefont {Karaiskaj}, \citenamefont {Bracker}, \citenamefont {Gammon},\
		and\ \citenamefont {Cundiff}}]{Moody2011}%
	\BibitemOpen
	\bibfield  {author} {\bibinfo {author} {\bibfnamefont {G.}~\bibnamefont
			{Moody}}, \bibinfo {author} {\bibfnamefont {M.~E.}\ \bibnamefont {Siemens}},
		\bibinfo {author} {\bibfnamefont {A.~D.}\ \bibnamefont {Bristow}}, \bibinfo
		{author} {\bibfnamefont {X.}~\bibnamefont {Dai}}, \bibinfo {author}
		{\bibfnamefont {D.}~\bibnamefont {Karaiskaj}}, \bibinfo {author}
		{\bibfnamefont {A.~S.}\ \bibnamefont {Bracker}}, \bibinfo {author}
		{\bibfnamefont {D.}~\bibnamefont {Gammon}},\ and\ \bibinfo {author}
		{\bibfnamefont {S.~T.}\ \bibnamefont {Cundiff}},\ }\bibfield  {title}
	{\enquote {\bibinfo {title} {Exciton-exciton and exciton-phonon interactions
				in an interfacial {GaAs} quantum dot ensemble},}\ }\href
	{https://doi.org/10.1103/PhysRevB.83.115324} {\bibfield  {journal} {\bibinfo
			{journal} {Phys. Rev. B}\ }\textbf {\bibinfo {volume} {83}},\ \bibinfo
		{pages} {115324} (\bibinfo {year} {2011})}\BibitemShut {NoStop}%
	\bibitem [{\citenamefont {Nardin}\ \emph {et~al.}(2014)\citenamefont {Nardin},
		\citenamefont {Moody}, \citenamefont {Singh}, \citenamefont {Autry},
		\citenamefont {Li}, \citenamefont {Morier-Genoud},\ and\ \citenamefont
		{Cundiff}}]{Nardin2014}%
	\BibitemOpen
	\bibfield  {author} {\bibinfo {author} {\bibfnamefont {G.}~\bibnamefont
			{Nardin}}, \bibinfo {author} {\bibfnamefont {G.}~\bibnamefont {Moody}},
		\bibinfo {author} {\bibfnamefont {R.}~\bibnamefont {Singh}}, \bibinfo
		{author} {\bibfnamefont {T.~M.}\ \bibnamefont {Autry}}, \bibinfo {author}
		{\bibfnamefont {H.}~\bibnamefont {Li}}, \bibinfo {author} {\bibfnamefont
			{F.~m.~c.}\ \bibnamefont {Morier-Genoud}},\ and\ \bibinfo {author}
		{\bibfnamefont {S.~T.}\ \bibnamefont {Cundiff}},\ }\bibfield  {title}
	{\enquote {\bibinfo {title} {Coherent excitonic coupling in an asymmetric
				double ingaas quantum well arises from many-body effects},}\ }\href
	{https://doi.org/10.1103/PhysRevLett.112.046402} {\bibfield  {journal}
		{\bibinfo  {journal} {Phys. Rev. Lett.}\ }\textbf {\bibinfo {volume} {112}},\
		\bibinfo {pages} {046402} (\bibinfo {year} {2014})}\BibitemShut {NoStop}%
	\bibitem [{\citenamefont {Moody}\ \emph {et~al.}(2015)\citenamefont {Moody},
		\citenamefont {Dass}, \citenamefont {Hao}, \citenamefont {Chen},
		\citenamefont {Li}, \citenamefont {Singh}, \citenamefont {Tran},
		\citenamefont {Clark}, \citenamefont {Xu}, \citenamefont {Bergh{\"a}user}
		\emph {et~al.}}]{moody2015intrinsic}%
	\BibitemOpen
	\bibfield  {author} {\bibinfo {author} {\bibfnamefont {G.}~\bibnamefont
			{Moody}}, \bibinfo {author} {\bibfnamefont {C.~K.}\ \bibnamefont {Dass}},
		\bibinfo {author} {\bibfnamefont {K.}~\bibnamefont {Hao}}, \bibinfo {author}
		{\bibfnamefont {C.-H.}\ \bibnamefont {Chen}}, \bibinfo {author}
		{\bibfnamefont {L.-J.}\ \bibnamefont {Li}}, \bibinfo {author} {\bibfnamefont
			{A.}~\bibnamefont {Singh}}, \bibinfo {author} {\bibfnamefont
			{K.}~\bibnamefont {Tran}}, \bibinfo {author} {\bibfnamefont {G.}~\bibnamefont
			{Clark}}, \bibinfo {author} {\bibfnamefont {X.}~\bibnamefont {Xu}}, \bibinfo
		{author} {\bibfnamefont {G.}~\bibnamefont {Bergh{\"a}user}}, \emph {et~al.},\
	}\bibfield  {title} {\enquote {\bibinfo {title} {Intrinsic homogeneous
				linewidth and broadening mechanisms of excitons in monolayer transition metal
				dichalcogenides},}\ }\href@noop {} {\bibfield  {journal} {\bibinfo  {journal}
			{Nat. Commun.}\ }\textbf {\bibinfo {volume} {6}},\ \bibinfo {pages} {8315}
		(\bibinfo {year} {2015})}\BibitemShut {NoStop}%
	\bibitem [{\citenamefont {Martin}\ \emph {et~al.}(2018)\citenamefont {Martin},
		\citenamefont {Horng}, \citenamefont {Ruth}, \citenamefont {Paik},
		\citenamefont {Wentzel}, \citenamefont {Deng},\ and\ \citenamefont
		{Cundiff}}]{martin2018encapsulation}%
	\BibitemOpen
	\bibfield  {author} {\bibinfo {author} {\bibfnamefont {E.~W.}\ \bibnamefont
			{Martin}}, \bibinfo {author} {\bibfnamefont {J.}~\bibnamefont {Horng}},
		\bibinfo {author} {\bibfnamefont {H.~G.}\ \bibnamefont {Ruth}}, \bibinfo
		{author} {\bibfnamefont {E.}~\bibnamefont {Paik}}, \bibinfo {author}
		{\bibfnamefont {M.-H.}\ \bibnamefont {Wentzel}}, \bibinfo {author}
		{\bibfnamefont {H.}~\bibnamefont {Deng}},\ and\ \bibinfo {author}
		{\bibfnamefont {S.~T.}\ \bibnamefont {Cundiff}},\ }\href@noop {} {\enquote
		{\bibinfo {title} {Encapsulation narrows excitonic homogeneous linewidth of
				exfoliated {MoSe$_2$} monolayer},}\ } (\bibinfo {year} {2018}),\ \bibinfo
	{note} {arXiv:1810.09834 [cond-mat.mtrl-sci]}\BibitemShut {NoStop}%
	\bibitem [{\citenamefont {Thouin}\ \emph
		{et~al.}(2019{\natexlab{a}})\citenamefont {Thouin}, \citenamefont
		{Cortecchia}, \citenamefont {Petrozza}, \citenamefont {Srimath~Kandada},\
		and\ \citenamefont {Silva}}]{thouin2019enhanced}%
	\BibitemOpen
	\bibfield  {author} {\bibinfo {author} {\bibfnamefont {F.}~\bibnamefont
			{Thouin}}, \bibinfo {author} {\bibfnamefont {D.}~\bibnamefont {Cortecchia}},
		\bibinfo {author} {\bibfnamefont {A.}~\bibnamefont {Petrozza}}, \bibinfo
		{author} {\bibfnamefont {A.~R.}\ \bibnamefont {Srimath~Kandada}},\ and\
		\bibinfo {author} {\bibfnamefont {C.}~\bibnamefont {Silva}},\ }\bibfield
	{title} {\enquote {\bibinfo {title} {Enhanced screening and spectral
				diversity in many-body elastic scattering of excitons in two-dimensional
				hybrid metal-halide perovskites},}\ }\href@noop {} {\bibfield  {journal}
		{\bibinfo  {journal} {Phys. Rev. Res.}\ }\textbf {\bibinfo {volume} {1}},\
		\bibinfo {pages} {032032} (\bibinfo {year} {2019}{\natexlab{a}})}\BibitemShut
	{NoStop}%
	\bibitem [{\citenamefont {Katsch}, \citenamefont {Selig},\ and\ \citenamefont
		{Knorr}(2020)}]{Katsch2020}%
	\BibitemOpen
	\bibfield  {author} {\bibinfo {author} {\bibfnamefont {F.}~\bibnamefont
			{Katsch}}, \bibinfo {author} {\bibfnamefont {M.}~\bibnamefont {Selig}},\ and\
		\bibinfo {author} {\bibfnamefont {A.}~\bibnamefont {Knorr}},\ }\bibfield
	{title} {\enquote {\bibinfo {title} {Exciton-scattering-induced dephasing in
				two-dimensional semiconductors},}\ }\href
	{https://doi.org/10.1103/PhysRevLett.124.257402} {\bibfield  {journal}
		{\bibinfo  {journal} {Phys. Rev. Lett.}\ }\textbf {\bibinfo {volume} {124}},\
		\bibinfo {pages} {257402} (\bibinfo {year} {2020})}\BibitemShut {NoStop}%
	\bibitem [{\citenamefont {Erkensten}, \citenamefont {Brem},\ and\ \citenamefont
		{Malic}()}]{Erkensten_EID_2020}%
	\BibitemOpen
	\bibfield  {author} {\bibinfo {author} {\bibfnamefont {D.}~\bibnamefont
			{Erkensten}}, \bibinfo {author} {\bibfnamefont {S.}~\bibnamefont {Brem}},\
		and\ \bibinfo {author} {\bibfnamefont {E.}~\bibnamefont {Malic}},\
	}\href@noop {} {\enquote {\bibinfo {title} {Excitation-induced dephasing in
				{2D} materials and {van der Waals} heterostructures},}\ }\bibinfo {note}
	{ArXiv:2006.08392 [cond-mat.mtrl-sci]}\BibitemShut {NoStop}%
	\bibitem [{\citenamefont {Srimath~Kandada}\ and\ \citenamefont
		{Silva}(2020)}]{SrimathKandada2020}%
	\BibitemOpen
	\bibfield  {author} {\bibinfo {author} {\bibfnamefont {A.~R.}\ \bibnamefont
			{Srimath~Kandada}}\ and\ \bibinfo {author} {\bibfnamefont {C.}~\bibnamefont
			{Silva}},\ }\bibfield  {title} {\enquote {\bibinfo {title} {Exciton polarons
				in two-dimensional hybrid metal-halide perovskites},}\ }\href
	{https://doi.org/10.1021/acs.jpclett.9b02342} {\bibfield  {journal} {\bibinfo
			{journal} {J. Phys. Chem. Lett.}\ }\textbf {\bibinfo {volume} {11}},\
		\bibinfo {pages} {3173--3184} (\bibinfo {year} {2020})}\BibitemShut {NoStop}%
	\bibitem [{\citenamefont {W.~Anderson}(1954)}]{w1954mathematical}%
	\BibitemOpen
	\bibfield  {author} {\bibinfo {author} {\bibfnamefont {P.}~\bibnamefont
			{W.~Anderson}},\ }\bibfield  {title} {\enquote {\bibinfo {title} {A
				mathematical model for the narrowing of spectral lines by exchange or
				motion},}\ }\href@noop {} {\bibfield  {journal} {\bibinfo  {journal} {J.
				Phys. Soc. Jpn.}\ }\textbf {\bibinfo {volume} {9}},\ \bibinfo {pages}
		{316--339} (\bibinfo {year} {1954})}\BibitemShut {NoStop}%
	\bibitem [{\citenamefont {Kubo}(1954)}]{kubo1954note}%
	\BibitemOpen
	\bibfield  {author} {\bibinfo {author} {\bibfnamefont {R.}~\bibnamefont
			{Kubo}},\ }\bibfield  {title} {\enquote {\bibinfo {title} {Note on the
				stochastic theory of resonance absorption},}\ }\href@noop {} {\bibfield
		{journal} {\bibinfo  {journal} {J. Phys. Soc. Jpn.}\ }\textbf {\bibinfo
			{volume} {9}},\ \bibinfo {pages} {935--944} (\bibinfo {year}
		{1954})}\BibitemShut {NoStop}%
	\bibitem [{\citenamefont {Siemens}\ \emph {et~al.}(2010)\citenamefont
		{Siemens}, \citenamefont {Moody}, \citenamefont {Li}, \citenamefont
		{Bristow},\ and\ \citenamefont {Cundiff}}]{siemens2010resonance}%
	\BibitemOpen
	\bibfield  {author} {\bibinfo {author} {\bibfnamefont {M.~E.}\ \bibnamefont
			{Siemens}}, \bibinfo {author} {\bibfnamefont {G.}~\bibnamefont {Moody}},
		\bibinfo {author} {\bibfnamefont {H.}~\bibnamefont {Li}}, \bibinfo {author}
		{\bibfnamefont {A.~D.}\ \bibnamefont {Bristow}},\ and\ \bibinfo {author}
		{\bibfnamefont {S.~T.}\ \bibnamefont {Cundiff}},\ }\bibfield  {title}
	{\enquote {\bibinfo {title} {Resonance lineshapes in two-dimensional
				{Fourier} transform spectroscopy},}\ }\href@noop {} {\bibfield  {journal}
		{\bibinfo  {journal} {Optics Express}\ }\textbf {\bibinfo {volume} {18}},\
		\bibinfo {pages} {17699--17708} (\bibinfo {year} {2010})}\BibitemShut
	{NoStop}%
	\bibitem [{\citenamefont {Bristow}\ \emph {et~al.}(2011)\citenamefont
		{Bristow}, \citenamefont {Zhang}, \citenamefont {Siemens}, \citenamefont
		{Cundiff},\ and\ \citenamefont {Mirin}}]{bristow2011separating}%
	\BibitemOpen
	\bibfield  {author} {\bibinfo {author} {\bibfnamefont {A.~D.}\ \bibnamefont
			{Bristow}}, \bibinfo {author} {\bibfnamefont {T.}~\bibnamefont {Zhang}},
		\bibinfo {author} {\bibfnamefont {M.~E.}\ \bibnamefont {Siemens}}, \bibinfo
		{author} {\bibfnamefont {S.~T.}\ \bibnamefont {Cundiff}},\ and\ \bibinfo
		{author} {\bibfnamefont {R.}~\bibnamefont {Mirin}},\ }\bibfield  {title}
	{\enquote {\bibinfo {title} {Separating homogeneous and inhomogeneous line
				widths of heavy-and light-hole excitons in weakly disordered semiconductor
				quantum wells},}\ }\href@noop {} {\bibfield  {journal} {\bibinfo  {journal}
			{J. Phys. Chem. B}\ }\textbf {\bibinfo {volume} {115}},\ \bibinfo {pages}
		{5365--5371} (\bibinfo {year} {2011})}\BibitemShut {NoStop}%
	\bibitem [{\citenamefont {Li}\ \emph {et~al.}()\citenamefont {Li},
		\citenamefont {{Srimath Kandada}}, \citenamefont {Silva},\ and\ \citenamefont
		{Bittner}}]{paper2}%
	\BibitemOpen
	\bibfield  {author} {\bibinfo {author} {\bibfnamefont {H.}~\bibnamefont
			{Li}}, \bibinfo {author} {\bibfnamefont {A.~R.}\ \bibnamefont {{Srimath
					Kandada}}}, \bibinfo {author} {\bibfnamefont {C.}~\bibnamefont {Silva}},\
		and\ \bibinfo {author} {\bibfnamefont {E.~R.}\ \bibnamefont {Bittner}},\
	}\href@noop {} {\enquote {\bibinfo {title} {Stochastic scattering theory for
				excitation induced dephasing: Comparison to the {Anderson-Kubo} lineshape},}\
	}\bibinfo {note} {ArXiv:2008.09218 [physics.chem-ph]}\BibitemShut {NoStop}%
	\bibitem [{\citenamefont {Neutzner}\ \emph {et~al.}(2018)\citenamefont
		{Neutzner}, \citenamefont {Thouin}, \citenamefont {Cortecchia}, \citenamefont
		{Petrozza}, \citenamefont {Silva},\ and\ \citenamefont
		{Srimath~Kandada}}]{Neutzner2018}%
	\BibitemOpen
	\bibfield  {author} {\bibinfo {author} {\bibfnamefont {S.}~\bibnamefont
			{Neutzner}}, \bibinfo {author} {\bibfnamefont {F.}~\bibnamefont {Thouin}},
		\bibinfo {author} {\bibfnamefont {D.}~\bibnamefont {Cortecchia}}, \bibinfo
		{author} {\bibfnamefont {A.}~\bibnamefont {Petrozza}}, \bibinfo {author}
		{\bibfnamefont {C.}~\bibnamefont {Silva}},\ and\ \bibinfo {author}
		{\bibfnamefont {A.~R.}\ \bibnamefont {Srimath~Kandada}},\ }\bibfield  {title}
	{\enquote {\bibinfo {title} {{Exciton-polaron spectral structures in two
					dimensional hybrid lead-halide perovskites}},}\ }\href
	{https://doi.org/10.1103/PhysRevMaterials.2.064605} {\bibfield  {journal}
		{\bibinfo  {journal} {Phys. Rev. Mater.}\ }\textbf {\bibinfo {volume} {2}},\
		\bibinfo {pages} {064605} (\bibinfo {year} {2018})}\BibitemShut {NoStop}%
	\bibitem [{\citenamefont {Thouin}\ \emph {et~al.}(2018)\citenamefont {Thouin},
		\citenamefont {Neutzner}, \citenamefont {Cortecchia}, \citenamefont
		{Dragomir}, \citenamefont {Soci}, \citenamefont {Salim}, \citenamefont {Lam},
		\citenamefont {Leonelli}, \citenamefont {Petrozza}, \citenamefont
		{Srimath~Kandada},\ and\ \citenamefont {Silva}}]{Thouin2018}%
	\BibitemOpen
	\bibfield  {author} {\bibinfo {author} {\bibfnamefont {F.}~\bibnamefont
			{Thouin}}, \bibinfo {author} {\bibfnamefont {S.}~\bibnamefont {Neutzner}},
		\bibinfo {author} {\bibfnamefont {D.}~\bibnamefont {Cortecchia}}, \bibinfo
		{author} {\bibfnamefont {V.~A.}\ \bibnamefont {Dragomir}}, \bibinfo {author}
		{\bibfnamefont {C.}~\bibnamefont {Soci}}, \bibinfo {author} {\bibfnamefont
			{T.}~\bibnamefont {Salim}}, \bibinfo {author} {\bibfnamefont {Y.~M.}\
			\bibnamefont {Lam}}, \bibinfo {author} {\bibfnamefont {R.}~\bibnamefont
			{Leonelli}}, \bibinfo {author} {\bibfnamefont {A.}~\bibnamefont {Petrozza}},
		\bibinfo {author} {\bibfnamefont {A.~R.}\ \bibnamefont {Srimath~Kandada}},\
		and\ \bibinfo {author} {\bibfnamefont {C.}~\bibnamefont {Silva}},\ }\bibfield
	{title} {\enquote {\bibinfo {title} {{Stable biexcitons in two-dimensional
					metal-halide perovskites with strong dynamic lattice disorder}},}\ }\href
	{https://doi.org/10.1103/PhysRevMaterials.2.034001} {\bibfield  {journal}
		{\bibinfo  {journal} {Phys. Rev. Mater.}\ }\textbf {\bibinfo {volume} {2}},\
		\bibinfo {pages} {034001} (\bibinfo {year} {2018})}\BibitemShut {NoStop}%
	\bibitem [{\citenamefont {Thouin}\ \emph
		{et~al.}(2019{\natexlab{b}})\citenamefont {Thouin}, \citenamefont
		{Valverde-Ch{\'a}vez}, \citenamefont {Quarti}, \citenamefont {Cortecchia},
		\citenamefont {Bargigia}, \citenamefont {Beljonne}, \citenamefont {Petrozza},
		\citenamefont {Silva},\ and\ \citenamefont
		{Srimath~Kandada}}]{thouin2019phonon}%
	\BibitemOpen
	\bibfield  {author} {\bibinfo {author} {\bibfnamefont {F.}~\bibnamefont
			{Thouin}}, \bibinfo {author} {\bibfnamefont {D.~A.}\ \bibnamefont
			{Valverde-Ch{\'a}vez}}, \bibinfo {author} {\bibfnamefont {C.}~\bibnamefont
			{Quarti}}, \bibinfo {author} {\bibfnamefont {D.}~\bibnamefont {Cortecchia}},
		\bibinfo {author} {\bibfnamefont {I.}~\bibnamefont {Bargigia}}, \bibinfo
		{author} {\bibfnamefont {D.}~\bibnamefont {Beljonne}}, \bibinfo {author}
		{\bibfnamefont {A.}~\bibnamefont {Petrozza}}, \bibinfo {author}
		{\bibfnamefont {C.}~\bibnamefont {Silva}},\ and\ \bibinfo {author}
		{\bibfnamefont {A.~R.}\ \bibnamefont {Srimath~Kandada}},\ }\bibfield  {title}
	{\enquote {\bibinfo {title} {Phonon coherences reveal the polaronic character
				of excitons in two-dimensional lead halide perovskites},}\ }\href@noop {}
	{\bibfield  {journal} {\bibinfo  {journal} {Nat. Mater.}\ }\textbf {\bibinfo
			{volume} {18}},\ \bibinfo {pages} {349--356} (\bibinfo {year}
		{2019}{\natexlab{b}})}\BibitemShut {NoStop}%
	\bibitem [{\citenamefont {Thouin}\ \emph
		{et~al.}(2019{\natexlab{c}})\citenamefont {Thouin}, \citenamefont
		{Srimath~Kandada}, \citenamefont {Valverde-Ch{\'a}vez}, \citenamefont
		{Cortecchia}, \citenamefont {Bargigia}, \citenamefont {Petrozza},
		\citenamefont {Yang}, \citenamefont {Bittner},\ and\ \citenamefont
		{Silva}}]{thouin2019polaron}%
	\BibitemOpen
	\bibfield  {author} {\bibinfo {author} {\bibfnamefont {F.}~\bibnamefont
			{Thouin}}, \bibinfo {author} {\bibfnamefont {A.~R.}\ \bibnamefont
			{Srimath~Kandada}}, \bibinfo {author} {\bibfnamefont {D.~A.}\ \bibnamefont
			{Valverde-Ch{\'a}vez}}, \bibinfo {author} {\bibfnamefont {D.}~\bibnamefont
			{Cortecchia}}, \bibinfo {author} {\bibfnamefont {I.}~\bibnamefont
			{Bargigia}}, \bibinfo {author} {\bibfnamefont {A.}~\bibnamefont {Petrozza}},
		\bibinfo {author} {\bibfnamefont {X.}~\bibnamefont {Yang}}, \bibinfo {author}
		{\bibfnamefont {E.~R.}\ \bibnamefont {Bittner}},\ and\ \bibinfo {author}
		{\bibfnamefont {C.}~\bibnamefont {Silva}},\ }\bibfield  {title} {\enquote
		{\bibinfo {title} {Electron-phonon couplings inherent in polarons drive
				exciton dynamics in two-dimensional metal-halide perovskites},}\ }\href
	{https://doi.org/10.1021/acs.chemmater.9b02267} {\bibfield  {journal}
		{\bibinfo  {journal} {Chem. Mater.}\ }\textbf {\bibinfo {volume} {31}},\
		\bibinfo {pages} {7085--7091} (\bibinfo {year}
		{2019}{\natexlab{c}})}\BibitemShut {NoStop}%
	\bibitem [{\citenamefont {Born}(1926)}]{born1926quantenmechanik}%
	\BibitemOpen
	\bibfield  {author} {\bibinfo {author} {\bibfnamefont {M.}~\bibnamefont
			{Born}},\ }\bibfield  {title} {\enquote {\bibinfo {title} {Quantenmechanik
				der sto{\ss}vorg{\"a}nge},}\ }\href@noop {} {\bibfield  {journal} {\bibinfo
			{journal} {Zeitschrift f{\"u}r Physik}\ }\textbf {\bibinfo {volume} {38}},\
		\bibinfo {pages} {803--827} (\bibinfo {year} {1926})}\BibitemShut {NoStop}%
	\bibitem [{\citenamefont {Steele}(2001)}]{Steele2001}%
	\BibitemOpen
	\bibfield  {author} {\bibinfo {author} {\bibfnamefont {J.~M.}\ \bibnamefont
			{Steele}},\ }\href {https://doi.org/10.1007/978-1-4684-9305-4} {\emph
		{\bibinfo {title} {Stochastic Calculus and Financial Applications}}}\
	(\bibinfo {year} {2001})\BibitemShut {NoStop}%
	\bibitem [{\citenamefont {von Weizs{\"{a}}cker}\ and\ \citenamefont
		{Winkler}(1990)}]{VonWeizsacker1990}%
	\BibitemOpen
	\bibfield  {author} {\bibinfo {author} {\bibfnamefont {H.}~\bibnamefont {von
				Weizs{\"{a}}cker}}\ and\ \bibinfo {author} {\bibfnamefont {G.}~\bibnamefont
			{Winkler}},\ }\bibfield  {title} {\enquote {\bibinfo {title}
			{{Ito-Calculus}},}\ \ }(\bibinfo {year} {1990})\BibitemShut {NoStop}%
	\bibitem [{\citenamefont {Fox}(1987)}]{Fox1987}%
	\BibitemOpen
	\bibfield  {author} {\bibinfo {author} {\bibfnamefont {R.~F.}\ \bibnamefont
			{Fox}},\ }\bibfield  {title} {\enquote {\bibinfo {title} {{Stochastic
					calculus in physics}},}\ }\href {https://doi.org/10.1007/BF01011160}
	{\bibfield  {journal} {\bibinfo  {journal} {J. Stat. Phys.}\ } (\bibinfo
		{year} {1987}),\ 10.1007/BF01011160}\BibitemShut {NoStop}%
	\bibitem [{\citenamefont {Mukamel}(1995)}]{Mukamel1995}%
	\BibitemOpen
	\bibfield  {author} {\bibinfo {author} {\bibfnamefont {S.}~\bibnamefont
			{Mukamel}},\ }\href@noop {} {\emph {\bibinfo {title} {{Principles of
					Nonlinear Optics and Spectroscopy}}}}\ (\bibinfo  {publisher} {Oxford
		University Press},\ \bibinfo {year} {1995})\BibitemShut {NoStop}%
	\bibitem [{\citenamefont {Kato}\ \emph {et~al.}(2003)\citenamefont {Kato},
		\citenamefont {Ichii}, \citenamefont {Ohashi}, \citenamefont {Kunugita},
		\citenamefont {Ema}, \citenamefont {Tanaka}, \citenamefont {Takahashi},\ and\
		\citenamefont {Kondo}}]{Kato2003}%
	\BibitemOpen
	\bibfield  {author} {\bibinfo {author} {\bibfnamefont {Y.}~\bibnamefont
			{Kato}}, \bibinfo {author} {\bibfnamefont {D.}~\bibnamefont {Ichii}},
		\bibinfo {author} {\bibfnamefont {K.}~\bibnamefont {Ohashi}}, \bibinfo
		{author} {\bibfnamefont {H.}~\bibnamefont {Kunugita}}, \bibinfo {author}
		{\bibfnamefont {K.}~\bibnamefont {Ema}}, \bibinfo {author} {\bibfnamefont
			{K.}~\bibnamefont {Tanaka}}, \bibinfo {author} {\bibfnamefont
			{T.}~\bibnamefont {Takahashi}},\ and\ \bibinfo {author} {\bibfnamefont
			{T.}~\bibnamefont {Kondo}},\ }\bibfield  {title} {\enquote {\bibinfo {title}
			{{Extremely large binding energy of biexcitons in an organic-inorganic
					quantum-well material \ce{C4H9NH3)2PbBr4}}},}\ }\href
	{https://doi.org/10.1016/S0038-1098(03)00626-4} {\bibfield  {journal}
		{\bibinfo  {journal} {Solid State Commun.}\ }\textbf {\bibinfo {volume}
			{128}},\ \bibinfo {pages} {15--18} (\bibinfo {year} {2003})}\BibitemShut
	{NoStop}%
	\bibitem [{\citenamefont {Cho}(2008)}]{cho2008coherent}%
	\BibitemOpen
	\bibfield  {author} {\bibinfo {author} {\bibfnamefont {M.}~\bibnamefont
			{Cho}},\ }\bibfield  {title} {\enquote {\bibinfo {title} {Coherent
				two-dimensional optical spectroscopy},}\ }\href@noop {} {\bibfield  {journal}
		{\bibinfo  {journal} {Chem. Rev.}\ }\textbf {\bibinfo {volume} {108}},\
		\bibinfo {pages} {1331--1418} (\bibinfo {year} {2008})}\BibitemShut {NoStop}%
	\bibitem [{\citenamefont {Tokmakoff}(2000)}]{tokmakoff2000two}%
	\BibitemOpen
	\bibfield  {author} {\bibinfo {author} {\bibfnamefont {A.}~\bibnamefont
			{Tokmakoff}},\ }\bibfield  {title} {\enquote {\bibinfo {title}
			{Two-dimensional line shapes derived from coherent third-order nonlinear
				spectroscopy},}\ }\href@noop {} {\bibfield  {journal} {\bibinfo  {journal}
			{J. Phys. Chem. A}\ }\textbf {\bibinfo {volume} {104}},\ \bibinfo {pages}
		{4247--4255} (\bibinfo {year} {2000})}\BibitemShut {NoStop}%
	\bibitem [{\citenamefont {Gauthron}\ \emph {et~al.}(2010)\citenamefont
		{Gauthron}, \citenamefont {Lauret}, \citenamefont {Doyennette}, \citenamefont
		{Lanty}, \citenamefont {Al~Choueiry}, \citenamefont {Zhang}, \citenamefont
		{Brehier}, \citenamefont {Largeau}, \citenamefont {Mauguin}, \citenamefont
		{Bloch} \emph {et~al.}}]{gauthron2010optical}%
	\BibitemOpen
	\bibfield  {author} {\bibinfo {author} {\bibfnamefont {K.}~\bibnamefont
			{Gauthron}}, \bibinfo {author} {\bibfnamefont {J.}~\bibnamefont {Lauret}},
		\bibinfo {author} {\bibfnamefont {L.}~\bibnamefont {Doyennette}}, \bibinfo
		{author} {\bibfnamefont {G.}~\bibnamefont {Lanty}}, \bibinfo {author}
		{\bibfnamefont {A.}~\bibnamefont {Al~Choueiry}}, \bibinfo {author}
		{\bibfnamefont {S.}~\bibnamefont {Zhang}}, \bibinfo {author} {\bibfnamefont
			{A.}~\bibnamefont {Brehier}}, \bibinfo {author} {\bibfnamefont
			{L.}~\bibnamefont {Largeau}}, \bibinfo {author} {\bibfnamefont
			{O.}~\bibnamefont {Mauguin}}, \bibinfo {author} {\bibfnamefont
			{J.}~\bibnamefont {Bloch}}, \emph {et~al.},\ }\bibfield  {title} {\enquote
		{\bibinfo {title} {Optical spectroscopy of two-dimensional layered
				{(C$_6$H$_5$C$_2$H$_4$-NH$_3$)$_2$-PbI$_4$} perovskite},}\ }\href@noop {}
	{\bibfield  {journal} {\bibinfo  {journal} {Optics Express}\ }\textbf
		{\bibinfo {volume} {18}},\ \bibinfo {pages} {5912--5919} (\bibinfo {year}
		{2010})}\BibitemShut {NoStop}%
	\bibitem [{\citenamefont {Tanaka}\ \emph {et~al.}(2002)\citenamefont {Tanaka},
		\citenamefont {Sano}, \citenamefont {Takahashi}, \citenamefont {Kondo},
		\citenamefont {Ito},\ and\ \citenamefont {Ema}}]{Tanaka2002}%
	\BibitemOpen
	\bibfield  {author} {\bibinfo {author} {\bibfnamefont {K.}~\bibnamefont
			{Tanaka}}, \bibinfo {author} {\bibfnamefont {F.}~\bibnamefont {Sano}},
		\bibinfo {author} {\bibfnamefont {T.}~\bibnamefont {Takahashi}}, \bibinfo
		{author} {\bibfnamefont {T.}~\bibnamefont {Kondo}}, \bibinfo {author}
		{\bibfnamefont {R.}~\bibnamefont {Ito}},\ and\ \bibinfo {author}
		{\bibfnamefont {K.}~\bibnamefont {Ema}},\ }\bibfield  {title} {\enquote
		{\bibinfo {title} {{Two-dimensional Wannier excitons in a
					layered-perovskite-type crystal \ce{(C6H13NH3)2PbI4}}},}\ }\href
	{https://doi.org/10.1016/S0038-1098(02)00126-6} {\bibfield  {journal}
		{\bibinfo  {journal} {Solid State Commun.}\ }\textbf {\bibinfo {volume}
			{122}},\ \bibinfo {pages} {249--252} (\bibinfo {year} {2002})}\BibitemShut
	{NoStop}%
	\bibitem [{\citenamefont {Tanaka}\ \emph {et~al.}(2005)\citenamefont {Tanaka},
		\citenamefont {Takahashi}, \citenamefont {Kondo}, \citenamefont {Umeda},
		\citenamefont {Ema}, \citenamefont {Umebayashi}, \citenamefont {Asai},
		\citenamefont {Uchida},\ and\ \citenamefont {Miura}}]{Tanaka2005}%
	\BibitemOpen
	\bibfield  {author} {\bibinfo {author} {\bibfnamefont {K.}~\bibnamefont
			{Tanaka}}, \bibinfo {author} {\bibfnamefont {T.}~\bibnamefont {Takahashi}},
		\bibinfo {author} {\bibfnamefont {T.}~\bibnamefont {Kondo}}, \bibinfo
		{author} {\bibfnamefont {K.}~\bibnamefont {Umeda}}, \bibinfo {author}
		{\bibfnamefont {K.}~\bibnamefont {Ema}}, \bibinfo {author} {\bibfnamefont
			{T.}~\bibnamefont {Umebayashi}}, \bibinfo {author} {\bibfnamefont
			{K.}~\bibnamefont {Asai}}, \bibinfo {author} {\bibfnamefont {K.}~\bibnamefont
			{Uchida}},\ and\ \bibinfo {author} {\bibfnamefont {N.}~\bibnamefont
			{Miura}},\ }\bibfield  {title} {\enquote {\bibinfo {title} {Electronic and
				excitonic structures of inorganic-organic perovskite-type quantum-well
				crystal \ce{(C4H9NH3)2PbBr4}},}\ }\href
	{https://doi.org/10.1143/JJAP.44.5923} {\bibfield  {journal} {\bibinfo
			{journal} {Japanese J. Appl. Physics, Part 1}\ }\textbf {\bibinfo {volume}
			{44}},\ \bibinfo {pages} {5923--5932} (\bibinfo {year} {2005})}\BibitemShut
	{NoStop}%
	\bibitem [{\citenamefont {Straus}\ and\ \citenamefont
		{Kagan}(2018)}]{Straus2018a}%
	\BibitemOpen
	\bibfield  {author} {\bibinfo {author} {\bibfnamefont {D.~B.}\ \bibnamefont
			{Straus}}\ and\ \bibinfo {author} {\bibfnamefont {C.~R.}\ \bibnamefont
			{Kagan}},\ }\bibfield  {title} {\enquote {\bibinfo {title} {Electrons,
				excitons, and phonons in two-dimensional hybrid perovskites: Connecting
				structural, optical, and electronic properties},}\ }\href
	{https://doi.org/10.1021/acs.jpclett.8b00201} {\bibfield  {journal} {\bibinfo
			{journal} {J. Phys. Chem. Lett.}\ }\textbf {\bibinfo {volume} {9}},\
		\bibinfo {pages} {1434--1447} (\bibinfo {year} {2018})}\BibitemShut {NoStop}%
	\bibitem [{\citenamefont {Mauck}\ and\ \citenamefont
		{Tisdale}(2019)}]{mauck2019excitons}%
	\BibitemOpen
	\bibfield  {author} {\bibinfo {author} {\bibfnamefont {C.~M.}\ \bibnamefont
			{Mauck}}\ and\ \bibinfo {author} {\bibfnamefont {W.~A.}\ \bibnamefont
			{Tisdale}},\ }\bibfield  {title} {\enquote {\bibinfo {title} {Excitons in
				{2D} organic--inorganic halide perovskites},}\ }\href@noop {} {\bibfield
		{journal} {\bibinfo  {journal} {Trends in Chemistry}\ } (\bibinfo {year}
		{2019})}\BibitemShut {NoStop}%
	\bibitem [{\citenamefont {Cortecchia}\ \emph {et~al.}(2016)\citenamefont
		{Cortecchia}, \citenamefont {Neutzner}, \citenamefont {Srimath~Kandada},
		\citenamefont {Mosconi}, \citenamefont {Meggiolaro}, \citenamefont
		{De~Angelis}, \citenamefont {Soci},\ and\ \citenamefont
		{Petrozza}}]{cortecchia2016broadband}%
	\BibitemOpen
	\bibfield  {author} {\bibinfo {author} {\bibfnamefont {D.}~\bibnamefont
			{Cortecchia}}, \bibinfo {author} {\bibfnamefont {S.}~\bibnamefont
			{Neutzner}}, \bibinfo {author} {\bibfnamefont {A.~R.}\ \bibnamefont
			{Srimath~Kandada}}, \bibinfo {author} {\bibfnamefont {E.}~\bibnamefont
			{Mosconi}}, \bibinfo {author} {\bibfnamefont {D.}~\bibnamefont {Meggiolaro}},
		\bibinfo {author} {\bibfnamefont {F.}~\bibnamefont {De~Angelis}}, \bibinfo
		{author} {\bibfnamefont {C.}~\bibnamefont {Soci}},\ and\ \bibinfo {author}
		{\bibfnamefont {A.}~\bibnamefont {Petrozza}},\ }\bibfield  {title} {\enquote
		{\bibinfo {title} {Broadband emission in two-dimensional hybrid perovskites:
				The role of structural deformation},}\ }\href@noop {} {\bibfield  {journal}
		{\bibinfo  {journal} {J. Am. Chem. Soc.}\ }\textbf {\bibinfo {volume}
			{139}},\ \bibinfo {pages} {39--42} (\bibinfo {year} {2016})}\BibitemShut
	{NoStop}%
	\bibitem [{\citenamefont {Tremblay}\ \emph {et~al.}(2019)\citenamefont
		{Tremblay}, \citenamefont {Thouin}, \citenamefont {Leisen}, \citenamefont
		{Bacsa}, \citenamefont {Srimath}, \citenamefont {Hoffman}, \citenamefont
		{Kanatzidis}, \citenamefont {Mohite}, \citenamefont {Silva}, \citenamefont
		{Barlow} \emph {et~al.}}]{tremblay20194npea}%
	\BibitemOpen
	\bibfield  {author} {\bibinfo {author} {\bibfnamefont {M.}~\bibnamefont
			{Tremblay}}, \bibinfo {author} {\bibfnamefont {F.}~\bibnamefont {Thouin}},
		\bibinfo {author} {\bibfnamefont {J.}~\bibnamefont {Leisen}}, \bibinfo
		{author} {\bibfnamefont {J.}~\bibnamefont {Bacsa}}, \bibinfo {author}
		{\bibfnamefont {K.~A.}\ \bibnamefont {Srimath}}, \bibinfo {author}
		{\bibfnamefont {J.}~\bibnamefont {Hoffman}}, \bibinfo {author} {\bibfnamefont
			{M.}~\bibnamefont {Kanatzidis}}, \bibinfo {author} {\bibfnamefont
			{A.}~\bibnamefont {Mohite}}, \bibinfo {author} {\bibfnamefont
			{C.}~\bibnamefont {Silva}}, \bibinfo {author} {\bibfnamefont
			{S.}~\bibnamefont {Barlow}}, \emph {et~al.},\ }\bibfield  {title} {\enquote
		{\bibinfo {title} {\ce{(4NPEA)2PbI4} ({4NPEA= 4-Nitrophenylethylammonium}):
				Structural, {NMR}, and optical properties of a 3$\times$ 3 corrugated {2D}
				hybrid perovskite.}}\ }\href@noop {} {\bibfield  {journal} {\bibinfo
			{journal} {J. Am. Chem. Soc.}\ }\textbf {\bibinfo {volume} {141}},\ \bibinfo
		{pages} {4521--4525} (\bibinfo {year} {2019})}\BibitemShut {NoStop}%
	\bibitem [{Note1()}]{Note1}%
	\BibitemOpen
	\bibinfo {note} {We note that $\gamma $, and not $2\gamma $, as defined in
		ref.\ 27, is the homegeneous linewidth obtained as a fit parameter from
		global analysis of the diagonal and antidiagonal cuts of the norm of the
		rephasing spectrum}\BibitemShut {NoStop}%
	\bibitem [{\citenamefont {Ema}\ \emph {et~al.}(2006)\citenamefont {Ema},
		\citenamefont {Umeda}, \citenamefont {Toda}, \citenamefont {Yajima},
		\citenamefont {Arai}, \citenamefont {Kunugita}, \citenamefont {Wolverson},\
		and\ \citenamefont {Davies}}]{Ema2006}%
	\BibitemOpen
	\bibfield  {author} {\bibinfo {author} {\bibfnamefont {K.}~\bibnamefont
			{Ema}}, \bibinfo {author} {\bibfnamefont {K.}~\bibnamefont {Umeda}}, \bibinfo
		{author} {\bibfnamefont {M.}~\bibnamefont {Toda}}, \bibinfo {author}
		{\bibfnamefont {C.}~\bibnamefont {Yajima}}, \bibinfo {author} {\bibfnamefont
			{Y.}~\bibnamefont {Arai}}, \bibinfo {author} {\bibfnamefont {H.}~\bibnamefont
			{Kunugita}}, \bibinfo {author} {\bibfnamefont {D.}~\bibnamefont
			{Wolverson}},\ and\ \bibinfo {author} {\bibfnamefont {J.~J.}\ \bibnamefont
			{Davies}},\ }\bibfield  {title} {\enquote {\bibinfo {title} {{Huge exchange
					energy and fine structure of excitons in an organic-inorganic quantum well
					material}},}\ }\href {https://doi.org/10.1103/PhysRevB.73.241310} {\bibfield
		{journal} {\bibinfo  {journal} {Phys. Rev. B}\ }\textbf {\bibinfo {volume}
			{73}},\ \bibinfo {pages} {241310(R)} (\bibinfo {year} {2006})}\BibitemShut
	{NoStop}%
	\bibitem [{\citenamefont {Kataoka}\ \emph {et~al.}(1993)\citenamefont
		{Kataoka}, \citenamefont {Kondo}, \citenamefont {Ito}, \citenamefont
		{Sasaki}, \citenamefont {Uchida},\ and\ \citenamefont
		{Miura}}]{kataoka1993magneto}%
	\BibitemOpen
	\bibfield  {author} {\bibinfo {author} {\bibfnamefont {T.}~\bibnamefont
			{Kataoka}}, \bibinfo {author} {\bibfnamefont {T.}~\bibnamefont {Kondo}},
		\bibinfo {author} {\bibfnamefont {R.}~\bibnamefont {Ito}}, \bibinfo {author}
		{\bibfnamefont {S.}~\bibnamefont {Sasaki}}, \bibinfo {author} {\bibfnamefont
			{K.}~\bibnamefont {Uchida}},\ and\ \bibinfo {author} {\bibfnamefont
			{N.}~\bibnamefont {Miura}},\ }\bibfield  {title} {\enquote {\bibinfo {title}
			{Magneto-optical study on excitonic spectra in \ce{(C6H13NH3)2PbI4}},}\
	}\href@noop {} {\bibfield  {journal} {\bibinfo  {journal} {Phys. Rev. B.}\
		}\textbf {\bibinfo {volume} {47}},\ \bibinfo {pages} {2010} (\bibinfo {year}
		{1993})}\BibitemShut {NoStop}%
	\bibitem [{\citenamefont {Urban}\ \emph {et~al.}(2020)\citenamefont {Urban},
		\citenamefont {Chehade}, \citenamefont {Dyksik}, \citenamefont {Menahem},
		\citenamefont {Surrente}, \citenamefont {Trippe-Allard}, \citenamefont
		{Maude}, \citenamefont {Garrot}, \citenamefont {Yaffe}, \citenamefont
		{Delporte} \emph {et~al.}}]{urban2020revealing}%
	\BibitemOpen
	\bibfield  {author} {\bibinfo {author} {\bibfnamefont {J.~M.}\ \bibnamefont
			{Urban}}, \bibinfo {author} {\bibfnamefont {G.}~\bibnamefont {Chehade}},
		\bibinfo {author} {\bibfnamefont {M.}~\bibnamefont {Dyksik}}, \bibinfo
		{author} {\bibfnamefont {M.}~\bibnamefont {Menahem}}, \bibinfo {author}
		{\bibfnamefont {A.}~\bibnamefont {Surrente}}, \bibinfo {author}
		{\bibfnamefont {G.}~\bibnamefont {Trippe-Allard}}, \bibinfo {author}
		{\bibfnamefont {D.~K.}\ \bibnamefont {Maude}}, \bibinfo {author}
		{\bibfnamefont {D.}~\bibnamefont {Garrot}}, \bibinfo {author} {\bibfnamefont
			{O.}~\bibnamefont {Yaffe}}, \bibinfo {author} {\bibfnamefont
			{E.}~\bibnamefont {Delporte}}, \emph {et~al.},\ }\bibfield  {title} {\enquote
		{\bibinfo {title} {Revealing excitonic phonon coupling in {(PE)$_2$(MA)$_{n-
						1}$Pb$_n$I$_{3n+1}$ 2D} layered perovskites},}\ }\href@noop {} {\bibfield
		{journal} {\bibinfo  {journal} {J. Phys. Chem. Lett.}\ } (\bibinfo {year}
		{2020})}\BibitemShut {NoStop}%
	\bibitem [{\citenamefont {Straus}\ \emph {et~al.}(2016)\citenamefont {Straus},
		\citenamefont {Hurtado~Parra}, \citenamefont {Iotov}, \citenamefont
		{Gebhardt}, \citenamefont {Rappe}, \citenamefont {Subotnik}, \citenamefont
		{Kikkawa},\ and\ \citenamefont {Kagan}}]{straus2016direct}%
	\BibitemOpen
	\bibfield  {author} {\bibinfo {author} {\bibfnamefont {D.~B.}\ \bibnamefont
			{Straus}}, \bibinfo {author} {\bibfnamefont {S.}~\bibnamefont
			{Hurtado~Parra}}, \bibinfo {author} {\bibfnamefont {N.}~\bibnamefont
			{Iotov}}, \bibinfo {author} {\bibfnamefont {J.}~\bibnamefont {Gebhardt}},
		\bibinfo {author} {\bibfnamefont {A.~M.}\ \bibnamefont {Rappe}}, \bibinfo
		{author} {\bibfnamefont {J.~E.}\ \bibnamefont {Subotnik}}, \bibinfo {author}
		{\bibfnamefont {J.~M.}\ \bibnamefont {Kikkawa}},\ and\ \bibinfo {author}
		{\bibfnamefont {C.~R.}\ \bibnamefont {Kagan}},\ }\bibfield  {title} {\enquote
		{\bibinfo {title} {Direct observation of electron--phonon coupling and slow
				vibrational relaxation in organic--inorganic hybrid perovskites},}\
	}\href@noop {} {\bibfield  {journal} {\bibinfo  {journal} {J. Am. Chem.
				Soc.}\ }\textbf {\bibinfo {volume} {138}},\ \bibinfo {pages} {13798--13801}
		(\bibinfo {year} {2016})}\BibitemShut {NoStop}%
	\bibitem [{\citenamefont {Roberts}, \citenamefont {Loparo},\ and\ \citenamefont
		{Tokmakoff}(2006)}]{roberts2006characterization}%
	\BibitemOpen
	\bibfield  {author} {\bibinfo {author} {\bibfnamefont {S.~T.}\ \bibnamefont
			{Roberts}}, \bibinfo {author} {\bibfnamefont {J.~J.}\ \bibnamefont
			{Loparo}},\ and\ \bibinfo {author} {\bibfnamefont {A.}~\bibnamefont
			{Tokmakoff}},\ }\bibfield  {title} {\enquote {\bibinfo {title}
			{Characterization of spectral diffusion from two-dimensional line shapes},}\
	}\href@noop {} {\bibfield  {journal} {\bibinfo  {journal} {J. Chem. Phys.}\
		}\textbf {\bibinfo {volume} {125}},\ \bibinfo {pages} {084502} (\bibinfo
		{year} {2006})}\BibitemShut {NoStop}%
	\bibitem [{\citenamefont {Tang}\ \emph {et~al.}(2020)\citenamefont {Tang},
		\citenamefont {Li}, \citenamefont {Li}, \citenamefont {Xu}, \citenamefont
		{Liu}, \citenamefont {Barmak}, \citenamefont {Watanabe}, \citenamefont
		{Taniguchi}, \citenamefont {MacDonald}, \citenamefont {Shan} \emph
		{et~al.}}]{tang2020simulation}%
	\BibitemOpen
	\bibfield  {author} {\bibinfo {author} {\bibfnamefont {Y.}~\bibnamefont
			{Tang}}, \bibinfo {author} {\bibfnamefont {L.}~\bibnamefont {Li}}, \bibinfo
		{author} {\bibfnamefont {T.}~\bibnamefont {Li}}, \bibinfo {author}
		{\bibfnamefont {Y.}~\bibnamefont {Xu}}, \bibinfo {author} {\bibfnamefont
			{S.}~\bibnamefont {Liu}}, \bibinfo {author} {\bibfnamefont {K.}~\bibnamefont
			{Barmak}}, \bibinfo {author} {\bibfnamefont {K.}~\bibnamefont {Watanabe}},
		\bibinfo {author} {\bibfnamefont {T.}~\bibnamefont {Taniguchi}}, \bibinfo
		{author} {\bibfnamefont {A.~H.}\ \bibnamefont {MacDonald}}, \bibinfo {author}
		{\bibfnamefont {J.}~\bibnamefont {Shan}}, \emph {et~al.},\ }\bibfield
	{title} {\enquote {\bibinfo {title} {Simulation of {Hubbard} model physics in
				\ce{WSe2}/\ce{W2} moir{\'e} superlattices},}\ }\href@noop {} {\bibfield
		{journal} {\bibinfo  {journal} {Nature}\ }\textbf {\bibinfo {volume} {579}},\
		\bibinfo {pages} {353--358} (\bibinfo {year} {2020})}\BibitemShut {NoStop}%
	\bibitem [{\citenamefont {Controzzi}, \citenamefont {Essler},\ and\
		\citenamefont {Tsvelik}(2001)}]{Controzzi2001}%
	\BibitemOpen
	\bibfield  {author} {\bibinfo {author} {\bibfnamefont {D.}~\bibnamefont
			{Controzzi}}, \bibinfo {author} {\bibfnamefont {F.~H.~L.}\ \bibnamefont
			{Essler}},\ and\ \bibinfo {author} {\bibfnamefont {A.~M.}\ \bibnamefont
			{Tsvelik}},\ }\bibfield  {title} {\enquote {\bibinfo {title} {Optical
				conductivity of one-dimensional mott insulators},}\ }\href
	{https://doi.org/10.1103/PhysRevLett.86.680} {\bibfield  {journal} {\bibinfo
			{journal} {Phys. Rev. Lett.}\ }\textbf {\bibinfo {volume} {86}},\ \bibinfo
		{pages} {680--683} (\bibinfo {year} {2001})}\BibitemShut {NoStop}%
	\bibitem [{\citenamefont {Schlappa}\ \emph {et~al.}(2012)\citenamefont
		{Schlappa}, \citenamefont {Wohlfeld}, \citenamefont {Zhou}, \citenamefont
		{Mourigal}, \citenamefont {Haverkort}, \citenamefont {Strocov}, \citenamefont
		{Hozoi}, \citenamefont {Monney}, \citenamefont {Nishimoto}, \citenamefont
		{Singh} \emph {et~al.}}]{schlappa2012spin}%
	\BibitemOpen
	\bibfield  {author} {\bibinfo {author} {\bibfnamefont {J.}~\bibnamefont
			{Schlappa}}, \bibinfo {author} {\bibfnamefont {K.}~\bibnamefont {Wohlfeld}},
		\bibinfo {author} {\bibfnamefont {K.}~\bibnamefont {Zhou}}, \bibinfo {author}
		{\bibfnamefont {M.}~\bibnamefont {Mourigal}}, \bibinfo {author}
		{\bibfnamefont {M.}~\bibnamefont {Haverkort}}, \bibinfo {author}
		{\bibfnamefont {V.}~\bibnamefont {Strocov}}, \bibinfo {author} {\bibfnamefont
			{L.}~\bibnamefont {Hozoi}}, \bibinfo {author} {\bibfnamefont
			{C.}~\bibnamefont {Monney}}, \bibinfo {author} {\bibfnamefont
			{S.}~\bibnamefont {Nishimoto}}, \bibinfo {author} {\bibfnamefont
			{S.}~\bibnamefont {Singh}}, \emph {et~al.},\ }\bibfield  {title} {\enquote
		{\bibinfo {title} {Spin--orbital separation in the quasi-one-dimensional
				{Mott} insulator \ce{Sr2CuO3}},}\ }\href@noop {} {\bibfield  {journal}
		{\bibinfo  {journal} {Nature}\ }\textbf {\bibinfo {volume} {485}},\ \bibinfo
		{pages} {82--85} (\bibinfo {year} {2012})}\BibitemShut {NoStop}%
	\bibitem [{\citenamefont {Turner}\ \emph {et~al.}(2011)\citenamefont {Turner},
		\citenamefont {Stone}, \citenamefont {Gundogdu},\ and\ \citenamefont
		{Nelson}}]{turner2011invited}%
	\BibitemOpen
	\bibfield  {author} {\bibinfo {author} {\bibfnamefont {D.~B.}\ \bibnamefont
			{Turner}}, \bibinfo {author} {\bibfnamefont {K.~W.}\ \bibnamefont {Stone}},
		\bibinfo {author} {\bibfnamefont {K.}~\bibnamefont {Gundogdu}},\ and\
		\bibinfo {author} {\bibfnamefont {K.~A.}\ \bibnamefont {Nelson}},\ }\bibfield
	{title} {\enquote {\bibinfo {title} {Invited article: The coherent optical
				laser beam recombination technique (colbert) spectrometer: Coherent
				multidimensional spectroscopy made easier},}\ }\href@noop {} {\bibfield
		{journal} {\bibinfo  {journal} {Rev. Sci. Instrum.}\ }\textbf {\bibinfo
			{volume} {82}},\ \bibinfo {pages} {081301} (\bibinfo {year}
		{2011})}\BibitemShut {NoStop}%
	\bibitem [{\citenamefont {Loriot}, \citenamefont {Gitzinger},\ and\
		\citenamefont {Forget}(2013)}]{loriot2013self}%
	\BibitemOpen
	\bibfield  {author} {\bibinfo {author} {\bibfnamefont {V.}~\bibnamefont
			{Loriot}}, \bibinfo {author} {\bibfnamefont {G.}~\bibnamefont {Gitzinger}},\
		and\ \bibinfo {author} {\bibfnamefont {N.}~\bibnamefont {Forget}},\
	}\bibfield  {title} {\enquote {\bibinfo {title} {Self-referenced
				characterization of femtosecond laser pulses by chirp scan},}\ }\href@noop {}
	{\bibfield  {journal} {\bibinfo  {journal} {Optics Express}\ }\textbf
		{\bibinfo {volume} {21}},\ \bibinfo {pages} {24879--24893} (\bibinfo {year}
		{2013})}\BibitemShut {NoStop}%
	\bibitem [{\citenamefont {Trebino}(2012)}]{trebino2012frequency}%
	\BibitemOpen
	\bibfield  {author} {\bibinfo {author} {\bibfnamefont {R.}~\bibnamefont
			{Trebino}},\ }\href@noop {} {\emph {\bibinfo {title} {Frequency-resolved
				optical gating: the measurement of ultrashort laser pulses}}}\ (\bibinfo
	{publisher} {Springer Science \& Business Media},\ \bibinfo {year}
	{2012})\BibitemShut {NoStop}%
\end{thebibliography}
%

%



\end{document}